\documentclass[10pt,prd,preprintnumbers,floatfix,aps,nofootinbib,notitlepage,showpacs]{revtex4} %twocolumn

\usepackage{latexsym}
\usepackage{epsfig}
\usepackage{amssymb}

\usepackage{endnotes}

\newcommand{\lp}{\left(}
\newcommand{\rp}{\right)}
\newcommand{\lb}{\left[}
\newcommand{\rb}{\right]}

\newcommand{\tr}{\mbox{tr}}

\newcommand{\ba}{\begin{eqnarray}}
\newcommand{\ea}{\end{eqnarray}}
\newcommand{\be}{\begin{equation}}
\newcommand{\ee}{\end{equation}}

\newcommand{\half}{{1\over 2}}

\newcommand{\al}{\alpha}
\newcommand{\bt}{\beta}
\newcommand{\ga}{\gamma}

\newcommand{\da}{\delta}
\newcommand{\la}{\lambda}

\newcommand{\Ga}{\Gamma}

\newcommand{\Lag}{\mathcal{L}}

\newcommand{\mc}{\mathcal}
\newcommand{\ph}{\phantom{\al}}

\newcommand{\ud}[2]{^{#1}_{\phantom{#1} #2}}
\newcommand{\du}[2]{_{#1}^{\phantom{#1} #2}}

\newcommand{\enangle}[1]{\langle #1 \rangle} %angle brackets
\newcommand{\enaangle}[1]{\langle\!\langle #1 \rangle\!\rangle}

\newcommand{\hashd}{\supset^{(!)}}

\usepackage{hyperref}

\begin{document}

\title{Transforming gravity: from derivative couplings to matter to second-order scalar-tensor theories beyond the Horndeski Lagrangian}

\author{Miguel Zumalac\'arregui$^{1,2}$}
\author{Juan Garc\'ia-Bellido$^{1}$}
\affiliation{$^{1}$Instituto de F\'isica Te\'orica IFT-UAM-CSIC, Universidad Aut\'onoma de Madrid,
C/ Nicol\'as Cabrera 13-15, Cantoblanco, 28049 Madrid, Spain}
\affiliation{$^2$Institut f\"ur Theoretische Physik, Ruprecht-Karls-Universit\"at Heidelberg,
Philosophenweg 16, 69120 Heidelberg, Germany}

\date{\today}


\begin{abstract}
We study the structure of scalar-tensor theories of gravity based on derivative couplings between the scalar and the matter degrees of freedom introduced through an effective metric.
%1) Clasification of such derivative interactions. Disformal transformations as general but appealing case (sufficiently restrictive, second order).
Such interactions are classified by their tensor structure into conformal (scalar), disformal (vector) and extended disformal (traceless tensor), as well as by the derivative order of the scalar field. Relations limited to first derivatives of the field ensure second order equations of motion in the Einstein frame and hence the absence of Ostrogradski ghost degrees of freedom. 
%2) Use the Jacobian to study the existence of an inverse mapping to the Jordan frame metric. Eigenvalues and Eigentensors. The path integral.
The existence of a mapping to the Jordan frame is not trivial in the general case, and can be addressed using the Jacobian of the frame transformation through its eigenvalues and eigentensors. These objects also appear in the study of different aspects of such theories, including the metric and field redefinition transformation of the path integral in the quantum mechanical description.
%3) Find a loophole in Horndeski's theorem.
Although sane in the Einstein frame, generic disformally coupled theories are described by higher order equations of motion in the Jordan frame.
This apparent contradiction is solved by the use of a hidden constraint: the contraction of the metric equations with a Jacobian eigentensor provides a constraint relation for the higher field derivatives, which allows one to express the dynamical equations in a second order form. This signals a loophole in Horndeski's theorem and allows one to enlarge the set of scalar-tensor theories which are Ostrogradski-stable.
%4) Gauss-Bonnet terms. 
The transformed Gauss-Bonnet terms are also discussed for the simplest conformal and disformal relations.
\end{abstract}

\maketitle

\section{Introduction}

Current cosmological observations agree on the fact that the universe is undergoing a late phase of accelerated expansion \cite{Perlmutter:1998np,Riess:1998cb,Hinshaw:2012aka,Anderson:2012sa,Ade:2013zuv}, analogous to the early-time, high-energy inflationary mechanism that is believed to have set the conditions necessary for Big-Bang cosmology \cite{1981PhRvD..23..347G}. The simplest explanation for such an acceleration in an otherwise matter dominated universe is provided by the inclusion of a cosmological constant, which is however very small compared to other energy scales present in the standard model of particle physics and which are expected to contribute to the universe's acceleration \cite{RevModPhys.61.1,Martin:2012bt}. This puzzle has triggered the revival and proposal of a number of alternative theories, which attempt to explain the surprising behavior of the universe on Hubble scales \cite{Copeland:2006wr,Clifton:2011jh}.

%Actions
Such theories are generally described by an action functional
 \begin{equation}
S[g_{\mu\nu},\psi^{(i)}] = \int d^4 x\,\sqrt{-g} \Lag[g_{\mu\nu},\psi^{(i)}]\,,   
 \end{equation}
in which the Lagrangian density $\Lag$ is a Lorentz-scalar which depends locally on the metric and matter fields ($g_{\mu\nu}$, $\psi^{(i)}$) and their derivatives. The classical dynamics followed by such fields are given by the Euler-Lagrange Equations
\begin{equation}\label{eq:euler_lagrange}
 \frac{\partial \Lag}{\partial \psi^{(i)}} - \nabla_\mu \frac{\partial\Lag}{\partial (\nabla_\mu \psi^{(i)})} 
+ \nabla_{\mu}\nabla_{\nu} \frac{\partial \Lag}{\partial (\nabla_\nu\nabla_\mu\psi^{(i)})} = 0\,,
\end{equation}
obtained by varying the action with respect to the fundamental fields. It has been further assumed that the action contains up to their second derivatives. 
%these fields have been defined on a (pseudo) Riemannian manifold

%Need for 2nd order Equations of motion \& Ostrogradski's theorem 
Any alternative theory has to fulfill a number of requirements in order to be satisfactory. A very strong limitation to the space of possible theories is given by \emph{Ostrogradski's theorem} \cite{Ostrogradski,Woodard:2006nt}: for a non-degenerate theory whose Lagrangian contains second or higher derivatives with respect to time, their associated Hamiltonian is unbounded from below, making the system unstable and lacking a well-defined vacuum state.
\emph{Degenerate theories} are those for which Ostrogradski's construction does not apply, as it is the case for any theory described by second order equations of motion. Such is the case of general relativity (GR), whose Lagrangian contains second derivatives of the metric, but with the right degenerate structure to be described by second order equations of motion. In fact, it is the only four dimensional, Lorentz-covariant local theory of a metric tensor which fulfills this requirement \cite{Lovelock:1971yv}.%
\footnote{Higher order terms are also acceptable if they represent perturbative corrections (e.g. to a low energy effective theory) \emph{and} this perturbative nature is enforced in the solutions \cite{Simon:1990ic}.}

When one considers scalar tensor theories of gravitation, \emph{Horndeski's theorem} \cite{Horndeski} determines the most general four dimensional, Lorentz-covariant, local scalar-tensor theory for which the variation (\ref{eq:euler_lagrange}) produces second order equations. It is described by a Lagrangian density of the form $\Lag_H=\Lag_2+\Lag_3+\Lag_4+\Lag_5$, with
\begin{eqnarray}
\Lag_2 &=&   G_2(X,\phi)\,, \label{LH2} 
 \\[5pt]
\Lag_3 &=& G_3(X,\phi) [\Phi] 
\,, \label{LH3}
 \\[5pt]
\Lag_4 &=& G_4(X,\phi) R + G_{4,X}\lp [\Phi]^2 - [\Phi^2] \rp\,, \label{LH4}
\\[5pt]
\Lag_5 &=& G_5(X,\phi) G_{\mu\nu}\phi^{;\mu\nu}  - \frac{1}{6}G_{5,X}\lp [\Phi]^3
- 3[\Phi][\Phi^2] + 2[\Phi^3] \rp \,, \label{LH5}
\end{eqnarray}
plus a matter Lagrangian (See \cite{Deffayet:2009mn,Deffayet:2011gz} for modern re-derivations).
The notation used in the above equations and throughout the article is presented in Table \ref{tab:notation}. The key to the degeneracy of the above theory is that second derivatives of the scalar field appear in anti-symmetric combinations, so that higher derivatives cancel in the Euler-Lagrange variation (\ref{eq:euler_lagrange}) if the free functions $G_3,G_4,G_5$ depend on the field derivatives through $X$.
A very important advantage of the Horndeski Lagrangian is that it contains many interesting physical theories (see \cite{Zumalacarregui:2012us} for a summary) and allows for a systematic study of their properties. Such a general approach has been applied to cosmological dynamics \cite{DeFelice:2011hq,Leon:2012mt}, compatibility with cosmological observations \cite{Baker:2012zs,Amendola:2012ky,Motta:2013cwa}, inflationary mechanisms \cite{Kobayashi:2011nu} and screening modifications of gravity \cite{Koyama:2013paa,Kase:2013uja} and the effective cosmological constant \cite{Charmousis:2011bf}. 
Besides General Relativity ($G_2=G_3=G_5=0$, $G_4=1/16\pi G$), the best known class of theories contained in $\Lag_H$ are Jordan-Brans-Dicke theories \cite{Brans:1961sx}, for which $G_3=G_5=0$, $G_4=f(\phi)/16\pi G$ and $G_2=X/\omega(\phi) - V(\phi)$. 

\begin{table*}
 \begin{tabular}{@{\quad} l @{\quad} } 
 \multicolumn{1}{c}{\quad Scalar Fields $\phi,\,\pi$}
\\ \hline\hline
$X=-\half g^{\mu\nu}\phi_{,\mu}\phi_{,\nu}$ $\to$ Canonical kinetic term for the scalar \\[3pt]
$\Phi_{\mu\nu}=\phi_{;\mu\nu}$, \
$\Phi^n_{\mu\nu}=\phi_{;\mu\alpha_1}\phi\ud{;\alpha_1}{;\alpha_2}\cdots\phi\ud{;\alpha_{n-1}}{;\nu}$ %(for $n>1$)
$\to$ Contraction of second derivatives of the scalar ($n$ fields) \\[3pt]
$[\Phi^n]=g^{\mu\nu}\Phi^n_{\mu\nu}$,  \ e.g.
 $[\Phi] = \phi\ud{;\mu}{;\mu}\equiv \Box\phi$, \ $[\Phi^2]= \phi_{;\al\bt}\phi^{;\al\bt}$...
$\to$ Traces with the metric  \\[2pt]  
$\enangle{\Phi^n}=\phi^{,\mu} \Phi^n_{\mu\nu}\phi^{,\nu}$,  \ e.g.
$\enangle{\Phi}=\phi_{,\al}\phi^{;\al\bt}\phi_{,\bt}$, \ $\enangle{\Phi^2}=\phi_{,\al}\phi^{;\al\la}\phi_{;\la\bt}\phi^{,\bt}$...
$\to$ Traces with the field derivatives \\[3pt]
\hline\hline \\[-5pt]
 \multicolumn{1}{c}{\quad Tensor Fields $g_{\mu\nu}\,, \bar g_{\mu\nu}\,, \tilde g_{\mu\nu}$...} \\[3pt]\hline\hline
{$g_{\mu\nu}$ $\to$ Dynamical metric, i.e. its dynamics determined by $\delta S/\delta g_{\mu\nu}=0$.} \\[2pt]
{$\bar g_{\mu\nu},\, \tilde g_{\mu\nu}$ $\to$ Effective metric with scalar field dependence, (cf. table \ref{table_beyondBekenstein})} \\[3pt]
$\nabla_\mu,\, \Gamma^{\al}_{\mu\nu}$ $\to$
Torsion-free covariant derivative and connection compatible with $g_{\mu\nu}$. Also $u\ud{\mu}{;\nu}=\nabla_\nu u^\mu$ \\[2pt]
$\bar \nabla_\mu,\, \bar \Gamma^{\al}_{\mu\nu}$, $\tilde \nabla_\mu,\, \tilde \Gamma^{\al}_{\mu\nu}$ $\to$ idem for $\bar g_{\mu\nu}$, $\tilde g_{\mu\nu}$ cf. Eq. (\ref{connections})
\textbf{- All barred/tilde quantities constructed out of $\bar g_{\mu\nu}$, $\tilde g_{\mu\nu}$}\\[2pt]
\hline\hline \\[-5pt]
 \multicolumn{1}{c}{\quad Curvature $R\ud{\al}{\bt\mu\nu}$ c.f. (\ref{riemmangen}} \\[3pt]\hline\hline
$[R_{\mu\nu}]=R_{\mu\nu}g^{\mu\nu}$, $\enangle{R_{\mu\nu}} = \phi^{,\nu}\phi^{,\mu}R_{\mu\nu}$,
$[R_{\mu\nu}^2]=R_{\mu\nu}R^{\mu\nu}$,
$\langle R_{\mu\nu}^2 \rangle =\phi^{,\al} R_{\al\mu}R^{\mu\bt}\phi_{,\bt}$... \\[2pt] 
$\langle R_{\mu\nu}R^{\al\mu\bt\nu}\rangle=R_{\mu\nu}R^{\al\mu\bt\nu}\phi_{,\al}\phi_{,\mu}$,
 $\enaangle{ R_{\mu\al\nu\bt}\Phi^{\al\la}\Phi^{\bt\sigma}}= \phi^{,\mu}\phi^{,\nu} R_{\mu\al\nu\bt}\Phi^{\al\la}\Phi^{\bt\sigma}\phi_{,\la}\phi_{,\sigma} $ 
 \\[2pt]
\hline\hline
 \end{tabular}
\caption{Notation used in the text. Quantities with a bar or a tilde are constructed using the barred or tilde metric. All metrics have the signature $(-,+,+,+)$ signature, and the Riemann tensor is defined by $ 2  \nabla_{[\mu}  \nabla_{\nu]} v^{\al} \equiv R^{\al}_{\;\bt\mu\nu} v^\bt $. Parenthesis/brackets between indices will denote symmetrization/antisymmetrization $t_{(\mu\nu)}=\half(t_{\mu\nu}+t_{\nu\mu})$, $t_{[\mu\nu]}=\half(t_{\mu\nu}-t_{\nu\mu})$. The symbol $\hashd$ will be used to denote the higher time derivatives of an expression.
The word \emph{frame} (or physical frame) refers to the set of variables on which the variation is performed (e.g. Einstein/Jordan frame). Units in which $c=1$ will be used throughout.}\label{tab:notation}
\end{table*}

%Old school scalar tensor and Jordan-Brans-Dicke and conformal Relation
An important aspect of Jordan-Brans-Dicke theories is that the coupling between the scalar field and the curvature, given by $G_4(\phi)$, can be eliminated by a \emph{conformal transformation} in which the metric is rescaled by a function of the field $g_{\mu\nu}\to G_4^{-1} g_{\mu\nu}$.
This allows one to obtain different representations of the same theory, usually known as \emph{frames}, depending on which variables are considered dynamical: The original formulation is known as the \emph{Jordan frame}: the field $\phi$ and the Ricci scalar couple directly, but the matter Lagrangian only involves the metric, with no direct interaction between $\phi$ and the matter degrees of freedom. Alternatively, one may perform the aforementioned conformal transformation to the \emph{Einstein frame} in which the gravitational Lagrangian has the Einstein-Hilbert form ($G_4=1/16\pi G$) but the matter sector is directly affected by the scalar field, which mediates an additional force. Both representations are equivalent at the classical level (cf. \cite{Flanagan:2004bz,Deruelle:2010ht,Chiba:2013mha} and references therein), and each of them offers useful insight into their characteristics and behavior.

%special disformal relation
A natural question is whether generalizations of the conformal relation can offer further insights into the general class of scalar-tensor theories given by Eqs. (\ref{LH2}-\ref{LH5}). This can be done in some special cases, the simplest of them being the Dirac-Born-Infeld Galileons, which describe induced gravity on 4D branes embedded in five-dimensional space \cite{deRham:2010eu}. For a quartic DBI Galileon with $G_4=\sqrt{1-2X/M^4}$, $G_5=0$, it is possible to eliminate the non-minimal coupling to the Ricci scalar by means of a 
\emph{special disformal transformation}
\begin{equation}\label{eq:disf_special}
 g_{\mu\nu}\to \tilde  g_{\mu\nu} = C(\phi) g_{\mu\nu} + D(\phi)\phi_{,\mu}\phi_{,\nu}\,.
\end{equation}
with $C=1,D=-1/M^4$ \cite{Zumalacarregui:2012us}. More generally, Horndeski's theory has been shown to be formally invariant under special disformal transformations, which amount to a redefinition of the free functions $G_2-G_5$ \cite{Bettoni:2013cba} . Therefore, it is natural to consider (\ref{eq:disf_special}) as an integral part of Horndeski's theory.

%derivative coupligns to matter
When the matter sector is considered, the Einstein frame representation of the DBI Galileons introduces a \emph{derivative coupling} between the scalar field and the matter degrees of freedom. This has important pheonomenological consequences, as it modifies the relative causal structure between $g_{\mu\nu}$ and $\tilde  g_{\mu\nu}$. Additionally, the derivative coupling allows for the \emph{disformal screening mechanism} \cite{Koivisto:2012za}, which can hide the scalar-mediated additional force in high density environments. This effect might be related to the Vainshtein screening mechanism \cite{Vainshtein:1972sx}, which hides the scalar force within a certain radius of point sources due to the non-linear derivative self interactions of the field caused by the degenerate terms (\ref{LH3}-\ref{LH5}), as both theories are classically equivalent \cite{Zumalacarregui:2012us}. 
%Other nice properties
The inclusion of derivative interaction also allows for \emph{shift symmetry}, i.e. invariance of the action under the transformation $\phi\to\phi + c$. Exact (or softly broken) shift symmetry can be used to prevent large contributions to the field mass term or interactions with matter arising from quantum corrections. Shift symmetry plays an important role in certain scalar-tensor theories, such as Higgs-dilaton cosmology \cite{GarciaBellido:2011de,Ferrara:2010yw,Ferrara:2010in}, and is a particular case of Galilean symmetry $\phi\to\phi + c + b_\mu x^\mu$, which provides further improvement on the quantum properties of the theory \cite{Nicolis:2008in} (for cosmological applications of Galileons see \cite{Barreira:2012kk,Barreira:2013jma}).
 
%more general disformal relations
Disformal relations were originally introduced by Bekenstein in a more general form in which $C$ and $D$ are also allowed to depend on $X$ \cite{Bekenstein:1992pj}
\begin{equation}\label{eq:disf_Bekenstein}
 \tilde  g_{\mu\nu}=C(X,\phi) g_{\mu\nu} + D(X,\phi) \phi_{,\mu}\phi_{,\nu}\,.
\end{equation} 
Such relations have turned out to be very fruitful in the construction of alternatives to General Relativity, to a large extent because they can distort the causal structure between the two space-times since the line elements are now related by $d\tilde  s^2 = C ds^2 + D(\phi_{,\mu}dx^\mu)^2$. Assuming $C>0$, a 4-vector that is null with respect to $g_{\mu\nu}$ will be space-like or time-like with respect to $\tilde  g_{\mu\nu}$ depending on whether $D$ is positive or negative locally. Applications of the disformal relation (\ref{eq:disf_Bekenstein}) include inflation \cite{Kaloper:2003yf}, varying speed of light theories \cite{Clayton:1998hv,Magueijo:2003gj,Bassett:2000wj}, gravitational alternatives to Dark Matter (DM) \cite{Bekenstein:1993fs,Bekenstein:2004ne,Milgrom:2009gv}, screening modifications of gravity \cite{Noller:2012sv,Koivisto:2012za,Zumalacarregui:2012us}, violation of Lorentz invariance \cite{Brax:2012hm}, massive gravity \cite{deRham:2010ik,deRham:2010kj}, Dark Energy \cite{Koivisto:2008ak,
Zumalacarregui:2010wj}, 
DM candidates from extra dimensions \cite{Cembranos:2003fu,Cembranos:2003mr} or string theory \cite{Koivisto:2013fta,KOIVISTO:2013jwa} and exotic DM interactions \cite{Bettoni:2011fs,Bettoni:2012xv}. Disformal relations also appear in generalized Palatini gravities \cite{Olmo:2009xy}, provide new symmetries of Maxwell's \cite{Goulart:2013laa} and Horndeski's \cite{Bettoni:2013cba} theories, allow for the Lorentzian signature of the metric to emerge dynamically \cite{Mukohyama:2013ew,Magueijo:2013yya} and might be used in the construction of renormalizable theories of gravity \cite{Mukohyama:2013gra}. Unlike conformal relations, disformal couplings have non-trivial effects on radiation and can affect photons, a possibility that has been studied in the context of laboratory tests \cite{Brax:2012ie} and cosmological implications \cite{vandeBruck:2012vq,vandeBruck:2013yxa,Brax:2013nsa}.

%the Problem
Could a disformal transformation be used to remove the derivative couplings between the scalar field and the curvature from the higher Horndeski's terms $\Lag_4,\,\Lag_5$? The fact that a disformal coupling to matter of the form (\ref{eq:disf_Bekenstein}) does not introduce second order terms in the dynamical equations suggests that the equivalent Jordan frame representation belongs to Horndeski's theory.
However,  Bettoni \& Liberati have shown that this is not the case \cite{Bettoni:2013cba}: the action of a general disformal transformation (\ref{eq:disf_Bekenstein}) on the gravitational sector generates terms that can not be expressed in the form (\ref{LH2}-\ref{LH5}) unless the transformation is of the special type (\ref{eq:disf_special}). 
One of the purposes of this work is to examine the apparent contradiction between the second order nature of the disformally 
coupled theory, versus its apparently higher order nature in the Jordan frame.
 
In section \ref{section:beyondBekenstein} we study generalizations of the disformal relation. These can be classified by their tensor structure into conformal (scalar), disformal (vector) and extended disformal (traceless tensor), as well as by the number of derivatives of the fields that take part in the transformation. For relations involving a scalar and a metric tensor field, Bekenstein's relation (\ref{eq:disf_Bekenstein}) turns out to be a fairly natural choice, for which only two free functions are allowed and the equations of motion contain at most second derivatives of the field. Terms constructed out of second field derivatives might be also considered. However, they allow the construction of infinitely many terms which would generically lead to higher order terms in the equations.

The existence of a Jordan frame is examined in section \ref{section:jacobian} by studying under which conditions it is possible to find an inverse mapping for the disformal relation. The existence of such an inverse transformation is not trivial in the general case, and can be determined by studying the determinant of the Jacobian associated with (\ref{eq:disf_Bekenstein}), seen as a function of $g_{\mu\nu}\to \tilde g_{\mu\nu}$: an inverse map exists around any point for which the Jacobian determinant is non-zero. The invertibility is studied in detail for a general conformal and disformal case using the eigenvalues and eigenvectors of the Jacobian, which are in turn related to other aspects of frame transformation, including the transformation properties of the path integral. Several examples of inverse mappings with and without singular points are discussed.

Having obtained the conditions for an inverse transformation to exist, we proceed to analyze the Jordan frame theory in section \ref{sect:frame_Jordan}, first in the case of a non-trivial conformal transformation and then for the general case, focusing on a gravitational sector which is of the Einstein-Hilbert form in the original frame. The terms generated by the transformation do not belong to Horndeski's theory and their variation leads to equations which contain up to fourth time derivatives of the field and third time derivatives of the metric. However, by contracting the metric equations with a Jacobian eigentensor, a relation is derived which can be used to remove the higher derivatives from the equations. The hidden, second order nature of disformally coupled theories in the Jordan frame signals a loophole in Horndenski's theorem, as its derivation does not take into account the possibility of using combinations of the (initially higher order) dynamical equations to construct a second order theory.

Section \ref{section:discuss} contains a discussion of the main results, open questions and possible applications of the methods developed. Appendix \ref{app:general_disf} contains equations that arise form the general disformal relation (\ref{eq:disf_Bekenstein}) and were to long to be included in the text. Appendix \ref{app:special_disf} presents certain equations for special disformal transformations, including the transformation rules for the Einstein-Hilbert and the Horndeski Lagrangians, as well as the Gauss-Bonnet terms in the pure special conformal and pure special disformal case.

\section{Derivative Couplings to Matter} \label{section:beyondBekenstein}

Let us start by examining some of the properties of the theories of gravity formulated in a frame in which the matter sector contains disformal couplings between the matter degrees of freedom, the gravitational metric and the scalar field. An immediate question is what extensions of the original disformal relation (\ref{eq:disf_Bekenstein}) can be proposed and whether they are physically viable.
We can classify the transformations of the metric according to two different criteria:
\begin{itemize}
 \item By the \emph{tensor structure}. In order to obtain a symmetric tensor, it is possible to consider a \emph{conformal} term proportional to the original metric, a \emph{disformal} term constructed out of a vector $d_\mu$ and an \emph{extended disformal} term consisting on a rank-two symmetric tensor $E_{(\mu\nu)}$:
 \begin{equation}\label{eq:disf_general}
  \tilde  g_{\mu\nu} = C g_{\mu\nu} + D d_\mu d_\nu + E_{(\mu\nu)}\,.
 \end{equation}
The first term preserves the causal structure associated to both metrics (i.e. null vectors are null with respect to the two metrics), while the second and third terms do not. The main difference between the non-conformal terms is that $d_\mu d_\nu$ introduces a privileged direction along $d_\mu$.%
\footnote{One may as well include several disformal terms $d_{\mu}^{(i)}d_{\nu}^{(i)}$, or even terms made out of spinors: 
$d_{\mu}=\bar\psi\gamma_\mu \psi,\, \bar\psi\gamma_\mu \gamma_5 \psi,\, \bar\psi\nabla_\mu \psi, \, \bar\psi \gamma_5 \nabla_\mu \psi$, $E_{(\mu\nu)}=\bar \psi \gamma_{\mu}\gamma_{\nu}\psi$ (here $\bar\psi = \psi^\dagger \ga^0$) as long as they are consistent with the parity and tensor structure of the metric.}
To eliminate the degeneracy associated with the extended disformal term, we must project away the former terms so that  $d^{\mu}d^{\nu}E_{(\mu\nu)}= g^{\mu\nu}E_{(\mu\nu)}=0$.
\item By the \emph{derivative order}, i.e. how many derivatives of the variables are allowed in the transformation. Here we will restrict to relations in which no derivatives of the metric are introduced, except through covariant derivatives.%
\footnote{An example of theories featuring two metrics whose relation involves derivatives are $\mc C$ and $\mc D$ theories, which allow a conformal dependence on $R$ and a disformal dependence with the tensor structure of $R_{\mu\nu}$ \cite{Amendola:2010bk}.} The highest derivatives allowed are important for the character of the dynamical equations describing the theory, which might become higher than second order if second derivatives are included. Furthermore, higher derivatives also provide further tensor structures, as they allow arbitrary contractions using the same objects.
\end{itemize}
Table \ref{table_beyondBekenstein} summarizes the result of this classification for metrics constructed out of a scalar field.

\begin{table*}
 \begin{tabular}{@{\hspace{.2cm}} c @{\hspace{20pt}}  c @{\hspace{5pt}} c @{\hspace{5pt}} c @{\hspace{5pt}}  @{\hspace{20pt}} c @{\hspace{.4cm}}}
 \hline\hline %\cline{2-5}\\ \cline{2-5}
	& \hspace{.1cm} Conformal & \multicolumn{2}{c  @{\hspace{5pt}}}{ \hspace{1cm} Disformal} & Dependence\\
	& 	&\hspace{.1cm}  Vector \hspace{.1cm} & Tensor &			of $C,D,E$ \\ 
\hline\hline %\cline{2-4}
 General	& $C g_{\mu\nu}$ & $D d_\mu d_\nu$ & $E_{(\mu\nu)}$ & $C, d^2, [E^n], d\!\cdot\! E^n \!\cdot\! d$ \\[1pt] 
\hline\hline %\cline{2-4}
$\phi_{\phantom{;\mu\nu}}$ 	& $Cg_{\mu\nu}$ &	$-$  &	$-$ & $\phi$ \\
$\phi_{,\mu\phantom{\nu}}$ & $Cg_{\mu\nu}$ & $D\phi_{,\mu}\phi_{,\nu}$ & $-$ & $\phi,X$ \\
$\phi_{;\mu\nu}$ & $Cg_{\mu\nu}$ 
& $D\phi_{,\mu}\phi_{,\nu} + \sum_{n,m}D_{m,n}\phi^{,\al}\Phi_{;\al(\mu}^{m}\Phi_{;\nu)\bt}^{n}\phi^{,\bt}$ 
\hspace{.1cm} &\hspace{.1cm} 
$E\phi_{;\mu\nu} + \sum_l E_l \Phi_{;\mu\nu}^l$ \hspace{.1cm}
& $\phi,X,[\Phi^n],\enangle{\Phi^{n}}$ \\[2pt] \hline\hline
$v_{\mu;\nu}$ & $Cg_{\mu\nu}$ 
& $Dv_\mu v_\nu  + \sum_{n,m}D_{m,n} v^{\al}v_{\al;(\mu}^{m}v_{\nu);\bt}^{n}v^{\bt}$  \hspace{.1cm} &\hspace{.1cm} 
$E v_{(\mu;\nu)} + \sum_l E_l v_{(\mu;\nu)}^l$ \hspace{.1cm} & $v^2, [v_{\mu\nu}^n], d\cdot v_{;\mu\nu}^n\cdot d$\\[2pt]\hline\hline
\end{tabular}
\caption{Possible relations between metrics, cf. Eq. (\ref{eq:disf_general}). The columns classify the possible tensor structures that can be considered in the transformation (middle columns) as well as the possible dependences of the free functions (last column), i.e. all the Lorentz-scalars that can be constructed out of the objects introduced. The last column indicates the possible dependences of the functions $C,D,E$. Here $\Phi^1_{\mu\nu}=\phi_{;\mu\nu}$ and 
$\Phi^n_{\mu\nu}=\phi_{;\mu\alpha_1}\phi^{;\alpha_1}_{\quad;\alpha_2}\cdots\phi^{;\alpha_{n-1}}_{\quad;\nu}$ for $n>1$ ($n$ indicates the number of twice differentiated fields).
The table considers both the general case and the case of scalar tensor theories in which zero, one or two field derivatives are allowed. Allowing second field derivatives in the disformal relation allows for an (in principle) arbitrary number of terms to be added, due to the possibility of constructing contractions of $\phi_{;\mu\nu}$ with free indices. The last row displays the possible terms arising from a vector field and its first derivatives.
\label{table_beyondBekenstein}
}
\end{table*}

%reasons to keep's Bekenstein's choice
In the original work, Bekenstein disregarded theories including higher than second derivatives of the scalar because he expected that such theories would lead to higher than second order equations of motion and unbounded Hamiltonians. 
In addition, Table \ref{table_beyondBekenstein} makes clear that the introduction of objects with two indices allows for a potentially infinite set of different contractions.%
\footnote{Some of these terms or their combinations might give rise to total derivatives, which do not contribute to the equations of motion. Note also that a field redefinition $\phi = \phi(Y,\pi)$ with $Y=-\half(\partial\pi^2)$ adds a vector-like, second derivative term to the standard disformal structure 
$\phi_{,\mu}\phi_{,\nu} = (\phi_{,Y})^2 \pi^{,\al}\pi^{,\bt}\pi_{;\al\mu}\pi_{;\bt\nu}
-2 (\phi_{,Y}\phi_{,\pi})\pi^{,\al}\pi_{,\al(\mu}\pi_{,\nu)} + (\phi_{,\pi})^2\pi_{,\mu}\pi_{,\nu}$.}
Finally, if (covariant) derivatives of non-scalar objects are considered, the relation generically introduces derivatives of the metric tensor through the Christoffel connection.
For the sake of simplicity according to the above considerations, we will restrict our attention to metrics constructed only out of first field derivatives, as in the original disformal relation (\ref{eq:disf_Bekenstein}).

%Algebraic requirements
Besides mapping $g_{\mu\nu}$ into another rank-two symmetric tensor $\tilde  g_{\mu\nu}$, other physical requirements are necessary for the disformal relation to be physically reasonable. In order to have a well defined inverse metric $\tilde g^{\mu\nu}$ and provide an invariant integration volume, $\tilde  g_{\mu\nu}$ needs to have non-vanishing determinant
\begin{equation} \label{eq:disf_det}
 \tilde  g =  C^3(C-2DX) g\neq 0
\end{equation}
(see appendix C of Ref. \cite{Bekenstein:2004ne} for a derivation of the above expression). 
The authors of Ref. \cite{Bruneton:2007si} suggest that the functions $C,D$ have to be chosen such that the previous condition is satisfied for all possible values of $X$, i.e. $C>0$ and $C>2DX$. However, it has been observed that the second condition is maintained dynamically in cosmological models, as the field slows down whenever $X$ approaches $C/(2D)$. This happens both in the case of disformal couplings to matter \cite{Koivisto:2012za,Zumalacarregui:2012us} and scalar field self coupling \cite{Zumalacarregui:2010wj} for $D=D(\phi)>0$, suggesting that it is not necessary to tailor the functions $C,D$.
%Differential requirements
Moreover, theories formulated in terms of $\tilde  g_{\mu\nu}$ should be required to have a well posed initial value problem and give rise to second order evolution equations.

\subsection{Matter-Scalar Interaction} \label{einsteinframe}

Theories in which the matter Lagrangian is formulated in terms of a tilde metric (\ref{eq:disf_Bekenstein}, \ref{eq:disf_general}) introduce interactions between the matter and scalar degrees of freedom.
If $S_m = \int d^4 x \sqrt{-\tilde  g}\tilde  \Lag_m(\tilde  g,\psi^{(m)})$, the invariance of $S_m$ under coordinate transformations $x^\mu\to x^\mu + \xi^\mu$ implies that $\delta S_m=0$ and hence 
\begin{equation}
\int d^4 x \sqrt{- \tilde  g}\lp \frac{1}{\sqrt{- \tilde  g}}\frac{\delta (\sqrt{-\tilde  g}\tilde  \Lag_m)}{\delta \tilde  g_{\mu\nu}} \delta_\xi \tilde  g_{\mu\nu} 
+ \frac{\delta \tilde  \Lag_m}{\delta \psi^{(m)}}\delta_\xi\psi^{(m)} \rp 
= \int d^4 x \sqrt{-\tilde  g} \lp \half \tilde  \nabla_\mu \tilde  T^{\mu\nu} \rp\xi_{\nu} = 0\,,
\end{equation}
where the coefficient of $\delta\tilde   g_{\mu\nu}$ is proportional to the energy momentum tensor defined with respect to the tilde metric,
the matter equations of motion $\delta \Lag_m /\delta \psi_i =0$ and the metric transformation $\delta_\xi \tilde  g_{\mu\nu} = \tilde  \nabla_{(\mu}\xi_{\mu)}$ \cite{Wald_book} have been used, and $\tilde  \nabla_\mu$ is a torsion-free covariant derivative compatible with $\tilde  g_{\mu\nu}$ (cf. Eq. (\ref{connections}) and appendix \ref{App:connection}).
Therefore, as a direct application of Noether's theorem, energy-momentum is covariantly conserved as long as $\tilde  g_{\mu\nu}$ is used consistently in the equations
\begin{equation}\label{eq:barredconservation}
 \tilde  \nabla_\mu \tilde  T^{\mu\nu} =0\,.
\end{equation}
This derivation is valid as long as the theory is invariant under coordinate transformations and the motion for $\psi_i$ are fully determined from $\delta \tilde  \Lag_m$ alone. Note that no assumptions about the form of the gravitational sector and/or the tilde metric have been made in the derivation.

For disformal couplings to matter containing the field and its first derivatives (\ref{eq:disf_Bekenstein}), the contribution of the matter Lagrangian to the scalar field equation reads
\begin{eqnarray}
\frac{1}{\sqrt{-\tilde  g}} \frac{\delta\sqrt{-\tilde  g}\Lag_m}{\delta \phi} &=&  -\tilde  T^{\mu\nu}\, \tilde \nabla_\mu\lp D \phi_{,\nu}\rp
 + \half \tilde \nabla_\al \lp \tilde  T^{\mu\nu}\phi^{\al}\lp C_{,X}g_{\mu\nu} + D_{,X}\phi_{,\mu}\phi_{,\nu}\rp\rp 
 + \lp C_{,\phi}g_{\mu\nu} + D_{,\phi}\phi_{,\mu}\phi_{,\nu}\rp \tilde  T^{\mu\nu}\,,
 \label{eq:Q_Bekenstein}
\end{eqnarray}
where the energy momentum tensor in the first term has ``escaped'' the derivative by virtue of tilde energy conservation (\ref{eq:barredconservation}). As a consequence of the chain rule $\tilde  T^{\mu\nu}$ appears contracted with the partial derivatives of the tilde metric, $\tilde  g_{\mu\nu,X}$ and $\tilde  g_{\mu\nu,\phi}$.
%special disformal
The above equation implies that {special disformal relations} (\ref{eq:disf_special}) do not introduce derivatives of $\tilde  T^{\mu\nu}$ in the field equations, as $C_{,X},D_{,X}=0$.
In this case, equations (\ref{eq:barredconservation}) and (\ref{eq:Q_Bekenstein}) are equivalent to the equations derived in Refs. \cite{Koivisto:2012za,Zumalacarregui:2012us} in terms of untilde quantities (using the appropriate connection (\ref{connectionlongX}) and contracting tilde matter conservation (\ref{eq:barredconservation}) with $\tilde  g_{\la\nu}$.). 
This simplification may also be  related to the fact that the Horndeski Lagrangian (\ref{LH2}-\ref{LH5}) is formally invariant under special disformal transformations (\ref{eq:disf_special}) \cite{Bettoni:2013cba}. 

Note that all the terms contributed by (\ref{eq:Q_Bekenstein}) to the field equations of motion are at most second order in field derivatives, and therefore do not introduce Ostrogradski instabilities. This does not generally hold if second field derivatives are allowed, as in the relation (\ref{eq:disf_general}) with the terms described in table \ref{table_beyondBekenstein}. The variation of the matter Lagrangian w.r.t. the scalar field then reads
\begin{equation}
\frac{2}{\sqrt{-\tilde  g}}\frac{\delta \sqrt{-\tilde  g}\tilde  \Lag_m}{\delta\phi} = 
\tilde  T^{\mu\nu}\frac{\partial \tilde  g_{\mu\nu}}{\partial\phi}
-  \tilde  \nabla_{\al}\lp \tilde  T^{\mu\nu}\frac{\partial \tilde  g_{\mu\nu}}{\partial\phi_{,\al}}\rp
+  \tilde \nabla_{\bt}\tilde \nabla_{\al}\lp \tilde  T^{\mu\nu}\frac{\partial \tilde  g_{\mu\nu}}{\partial\phi_{;\al\bt}}\rp 
- \tilde \nabla_\la \lp \tilde  T^{\mu\nu}\frac{\partial \tilde  g_{\mu\nu}}{\partial\phi_{;\al\bt}} \mc K\ud{\la}{\al\bt} \rp \,,
\end{equation}
where $\mc K\ud{\la}{\al\bt}\equiv \tilde \Gamma\ud{\la}{\al\bt}-\Gamma\ud{\la}{\al\bt}$ is the difference between the connections for the field dependent and dynamical metrics, as given by Eq. (\ref{connections}) below. Even if it is possible to choose the coefficients of the disformal relation shown in table \ref{table_beyondBekenstein} to achieve second order equations of motion (including the difference between the connections $\mc K\ud{\la}{\al\bt}$), second derivatives of the field also introduce second derivatives of the energy momentum tensor. Although such theories are not a priory ruled out, they can lead to phenomenological problems (see Ref. \cite{Pani:2013qfa} for a discussion of such terms in the gravitational equations).

\section{Frame Transformations: The Jacobian}\label{section:jacobian}

Let us now study under which conditions disformally coupled theories can be re-expressed in the \emph{Jordan frame}, i.e. using the metric to which matter couples minimally as a fundamental variable. Ultimately, finding the {Jordan frame} frame requires inverting the relation between the two metrics and transforming the gravitational and matter sector accordingly. In the most thoroughly explored case of a special disformal relations (\ref{eq:disf_special}), such an inverse can be obtained trivially
\begin{equation}\label{eq:inv_disf_special}
g_{\mu\nu} = \frac{1}{C} \tilde g_{\mu\nu} - \frac{D}{C} \phi_{,\mu}\phi_{,\nu}\,,
\end{equation}
and is well defined as long as  $C\neq0$.

In more general cases, it is possible to address the existence of inverse map between the metrics by using the \emph{inverse function theorem} \cite{spivak-calc-manif}. Given a continuous differentiable function (that may depend on several variables), the theorem ensures the existence of an inverse, continuous differentiable function in a neighborhood of a point whenever the Jacobian determinant of the transformation is different from zero at that point. When applied to a function $\tilde g_{\mu\nu}(  g_{\al\bt})$, it implies that an inverse $  g_{\mu\nu}(\tilde g_{\al\bt})$ exists around any point for which
\begin{equation}\label{inverse_condition}
 \left|\frac{\partial \tilde g_{\mu\nu}}{\partial   g_{\al\bt}}\right|\neq 0\,,
\end{equation}
where we think of $\partial \tilde g_{\mu\nu}/\partial   g_{\al\bt}$ as a linear mapping of (0,2) symmetric tensors into (0,2) symmetric tensors (antisymmetric tensors are mapped to zero due to index symmetry).
{Note that the condition (\ref{inverse_condition}) only ensures the existence of an inverse map for the covariant metric. The existence of a contravariant metric $\tilde g^{\mu\nu}$ satisfying $\tilde g^{\al\la}\tilde g_{\la\bt}=\delta^\al_\bt$ requires that $\tilde g_{\mu\nu}$ has non-vanishing determinant, cf. (\ref{eq:disf_det}). This issue may be addressed by studying the \emph{contravariant Jacobian} $\partial \tilde g^{\mu\nu}/\partial g^{\al\bt}$. Demanding the absence of zeros might provide additional conditions for the frame transformation to exist, but these will not be considered here.
The theorem also determines that the Jacobian determinants of both mappings are the inverse of each other, 
$ \left| {\partial \tilde g_{\mu\nu}}/{\partial g_{\al\bt}} \right| = 
 \left| {\partial  g_{\mu\nu}}/{\partial \tilde g_{\al\bt}} \right|^{-1}$.}

The Jacobian determinant can be evaluated in terms of the eigenvalues of the Jacobian  $\left|\partial \tilde g_{\mu\nu}/\partial g_{\al\bt}\right| = \prod_n \la_n$, where each eigenvalue satisfies
\begin{equation}\label{eigenvalues}
\lp \frac{\partial \tilde g_{\mu\nu}}{\partial g_{\al\bt}} - \la_i \delta^\al_\mu\delta^\bt_\nu\rp \xi^{(i)}_{\al\bt}=0\,,
\end{equation}
for its associated eigentensor $\xi^{(i)}_{\al\bt}$. It is easy to check that for the special disformal transformation the only eigenvalue is $\la_C=C$, as expected from the explicit inverse metric (\ref{eq:inv_disf_special}). The Jacobian and and its eigenvalues will be studied in the non-trivial conformal case and for a general disformal metric (\ref{eq:disf_Bekenstein}) in the following subsections. However, the formalism presented above is general as long as the transformations of the metric tensor depends on the metric algebraically, and could be applied to more general relations such as (\ref{eq:disf_general}).%
\footnote{If derivatives of $\bar g_{\mu\nu}$ are introduced in the relation, the Jacobian may be generalized to
\begin{equation}
\frac{\delta \tilde g_{\mu\nu}}{\delta g_{\al\bt}} = \frac{\partial \tilde g_{\mu\nu}}{\partial g_{\al\bt}}
-\partial_\la\frac{\partial \tilde g_{\mu\nu}}{\partial g_{\al\bt,\la}}\,,
\end{equation}
but in this case it can not be used to determine the existence of an inverse map, which would be given by a differential equation.}

Besides determining the existence of an inverse transformation, the Jacobian occurs naturally at different points in the analysis of theories which can be formulated in different frames:
\begin{itemize}
\item \textbf{Energy-Momentum Tensor in different frames.}
The Jacobian  $\frac{\partial \tilde g_{\mu\nu}}{\partial   g_{\al\bt}}$ determines the relationship of the energy momentum tensor in different frames, which are related by the associated transformation
\begin{equation}\label{T_transform}
   T^{\mu\nu}=\sqrt{\frac{\tilde g}{ g}}  \frac{\partial \tilde g_{\al\bt}}{\partial g_{\mu\nu}} 
\tilde T^{\al\bt}\,,
\end{equation}
where $T^{\mu\nu}$ and $\tilde T^{\al\bt}$ are defined as the variation of the matter action with respect to the untilde and tilde metric respectively, as in Eq. (\ref{eq:Q_Bekenstein}). The energy momentum tensor obtained from the matter metric has the usual interpretation in terms of the energy fluxes seen by observers, while the one obtained with respect to the dynamical metric sources gravitational equations. The Jacobian  can be used to analyze the relation between them.

\item\textbf{Dynamical equations in the Jordan frame.}
The Jacobian also appears in the equations of motion in the Jordan frame through the chain rule
\begin{equation}\label{jacobian_variation}
\delta S_G = \frac{\delta S_G}{\delta \tilde g_{\al\bt}}\frac{\partial \tilde g_{\al\bt}}{\partial g_{\mu\nu}}\delta g_{\mu\nu} +\cdots \,,
\end{equation}
where the remaining terms in the variation have been omitted. This will be studied in detail in Section \ref{sect:frame_Jordan}, where the Jacobian will be used to re-write the theory in terms of second order equations of motion.

\item\textbf{Quantum mechanical formulation.}
The variational principle (\ref{eq:euler_lagrange}) obeyed by classical systems can be understood as a consequence of quantum mechanics in terms of the path integral
\begin{equation}
 Z(J) = \int \mc D \mc Q_i \exp\lp{-\frac{i}{h}\lp S[\mc Q_i] + \int d^4 x J^i \mc Q_i\rp}\rp\,,
\end{equation}
where $J^i$ is an external source and the integration is performed over all possible configurations of the dynamical variables, collectively denoted $\mc Q_i$.
In this interpretation, the imaginary exponent weights the probability associated to any given process. For configurations away from the classical solution the integrand oscillates rapidly and the amplitudes interfere destructively. As the classical solution minimizes the exponent, it will provide the only non-zero probability in the limit $h\to 0$.

If one considers an initial theory with a given set of fundamental variables and wishes to change the physical frame, e.g. from $\tilde g_{\mu\nu},\phi$ to $   g_{\mu\nu},\phi$, then the integration element in field space transforms as
\begin{equation}
 \mc D \mc Q_i = \mc D \phi \mc D \tilde g_{\mu\nu}
 \to  \mc D \phi \mc D   g_{\al\bt} \left|\frac{\partial \tilde g_{\mu\nu}}{\partial   g_{\al\bt}}\right|\,.
\end{equation}
By using the relation $\det M = \exp(\tr(\log(M)))$ it is possible to argue that the classical action picks up extra terms from the Jacobian. This is analogous to the occurrence of quantum anomalies, which typically lead to total derivatives and do not contribute to the classical equations of motion (e.g. \cite{Fujikawa:1980vr}). This is in agreement with the classical, but not necessarily quantum, equivalence between frames (see also Refs. \cite{Flanagan:2004bz,Fujii:1989qk}).

The transformation properties of the path integral are hence essentially linked to the problem of equivalence between physical frames. However, the quantum mechanical formulation of derivatively coupled scalar-tensor theories lies beyond the scope of this work, and in what follows we will consider all fields as classical and all frames as physically equivalent. The consequences of non-trivial metric transformations for quantum mechanics will be addressed elsewhere.
\end{itemize}

\subsection{Derivative Conformal Relation} \label{inverse_C_X}

%existence of an inverse
For a conformal relation depending on the field derivatives, $\tilde g_{\mu\nu}=C(X,\phi) g_{\mu\nu}$, the Jacobian  reads
\begin{equation}
\frac{\partial \tilde g_{\mu\nu}}{\partial  g_{\al\bt}} = C\delta^\al_\mu\delta^\bt_\nu + \half C_{,X} g_{\mu\nu}\phi^{,\al}\phi^{\bt}\,.
\end{equation}
As the dependence on $\phi$ does not alter the form of the equations, it will be omitted from them in the following. The equation for the eigenvalues (\ref{eigenvalues}) then follows
\begin{equation}
 (C-\la)\xi_{\mu\nu} + \half C_{,X} g_{\mu\nu} \enangle{\xi_{\al\bt}} =0\,,
\end{equation}
with $\enangle{\xi_{\al\bt}}\equiv \phi^{,\al}\xi_{\al\bt}\phi^{,\bt}$. The set of eigenvalues and eigentensors can be readily found from the above relation:
\begin{eqnarray}\label{conf_eigen1}
\la_C=C,\phantom{- C_{,X}X\, } &\;&  \xi^{C}_{\mu\nu} = v^{(1)}_{(\mu}v^{(2)}_{\nu)},
\quad 
\text{with } v^{(n)}_\mu\phi^{,\mu}=0
\,, \\
\la_K=  C - C_{,X}X,  && \xi^K_{\mu\nu} =  C_{,X} g_{\mu\nu} \label{conf_eigen2}
\end{eqnarray}
The \emph{conformal eigenvalue} (\ref{conf_eigen1}) was already found in the discussion of the special disformal relations (\ref{eq:disf_special}) and their inverse (\ref{eq:inv_disf_special}). It 
is degenerate with multiplicity 9, as there are $3^2$ directions orthogonal to $\phi_{,\nu}$. 
The new, characteristic feature of conformal derivative couplings comes from the \emph{kinetic eigenvalue} (\ref{conf_eigen2}), which is associated with the non-trivial dependence on $X$. It is non-degenerate with multiplicity 1, and becomes equal to $\la_C$ whenever $C_X$ or $X$ are zero (the normalization has been chosen for consistency with the general case studied in the next section). Note that there is no eigentensor proportional to $\phi_{,\mu}\phi_{\nu}$, as this is no privileged direction for $\tilde g_{\mu\nu}$. 

%relation to T_ab 
The kinetic eigenvalue (\ref{conf_eigen2}) has an associated eigentensor proportional to $g_{\mu\nu}$. This has important implications for the relation between the energy-momentum tensors defined with respect to the tilde and untilde metric, as given by Eq. (\ref{T_transform}), and in particular for their traces, which fulfill
\begin{equation} \label{conftrace}
 g_{\mu\nu}T^{\mu\nu} \propto (C-C_{,X}X)  \tilde g_{\mu\nu} \tilde T^{\mu\nu}\,.
\end{equation}
This type of relation might be used to analyze energy conditions in derivatively coupled scalar-tensor theories, analogously to similar studies in other alternative theories of gravity (e.g. \cite{Albareti:2012va}).

\begin{figure*}[t]
 \includegraphics[width=0.47\textwidth]{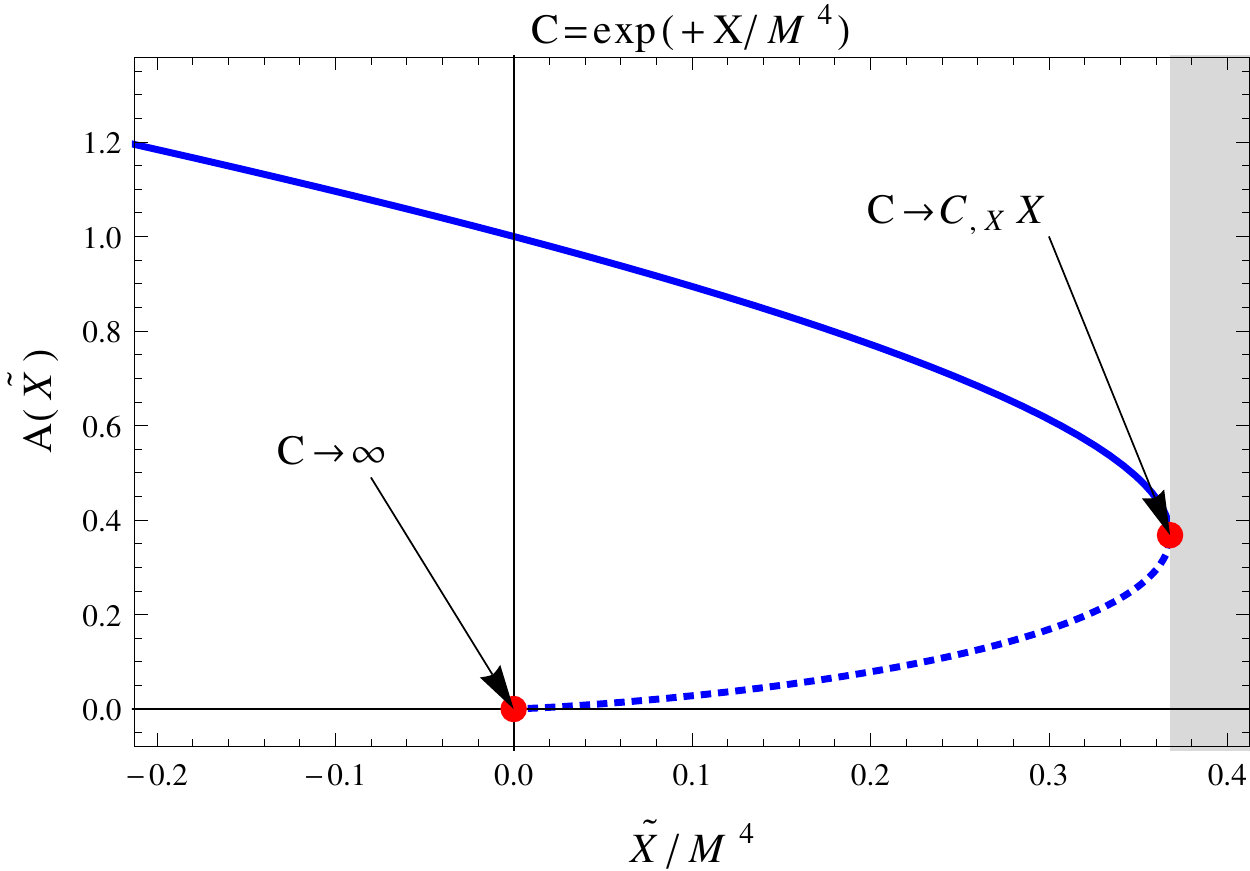} \hfill
 \includegraphics[width=0.47\textwidth]{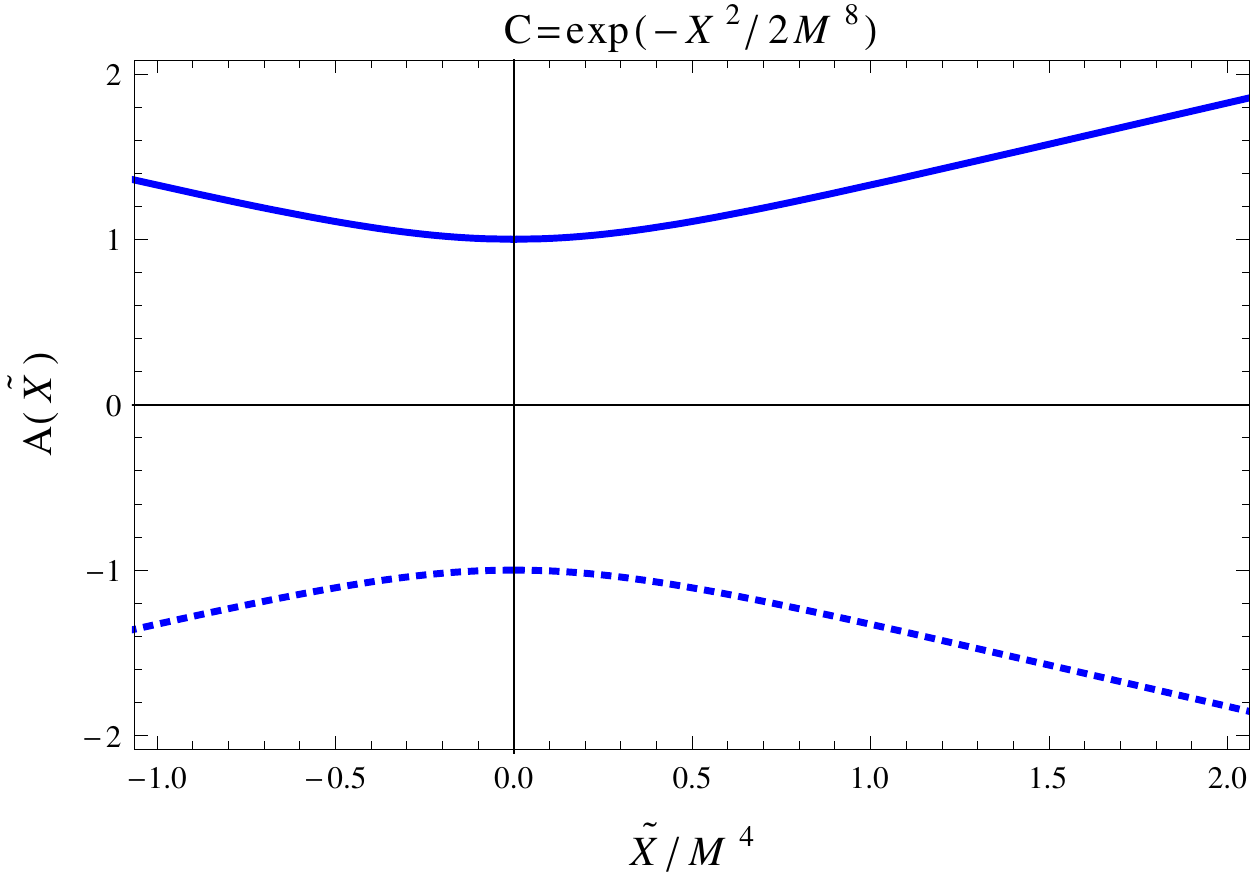}\hspace{2pt}
\caption{Inverse transformation $g_{\mu\nu}=A(\tilde X)\tilde g_{\mu\nu}$ for $\tilde g_{\mu\nu}=e^{X/M^4}g_{\mu\nu}$ (left) and $\tilde g_{\mu\nu}=e^{-X^2/(2M^8)}g_{\mu\nu}$ (right). The inverse conformal factor $A(\tilde X)$ is obtained implicitly through Eq. (\ref{inv_conf1}), and can be multi-valuated for certain ranges of $\tilde X$, giving rise to two branches characterized by the kinetic eigenvalue $\la_K$, given by Eq. (\ref{conf_eigen2}). The branches with positive (solid blue) and the negative (dotted blue) values of $\la_K$ meet or end at singular points, in which either one of the Jacobian eigenvalues (\ref{conf_eigen1}, \ref{conf_eigen2}) vanishes. 
This is seen explicitly in the case of the exponential function (left), which becomes bi-valued for $\tilde X/M^4>0$. Both branches meet at $\tilde X/M^4 = e^{-1}$ (corresponding to $X/M^4=1$) for which the kinetic eigenvalue becomes zero. The singular point $A(\tilde X)=0$ corresponds to $C(X)\to \infty$. 
The gray shaded region, $\tilde X/M^4 \ge e^{-1}$, is forbidden.
In the Gaussian case (right) both eigenvalues are always positive. Therefore there are no singular points and the two branches are not connected.
\label{fig_inverseConfExp}}
\end{figure*}

%form of the inverse
The inverse transformation $\tilde g_{\mu\nu}(g_{\mu\nu})$ is not well defined around points for which either $C$ or $C-C_{,X}X$ are equal to zero. Wherever the inverse exist, it will be of the conformal form in order for both metrics to represent the same space-time, with unaltered null-geodesics. We can hence use  $g_{\mu\nu} = A(\tilde X) \tilde g_{\mu\nu}$, $g^{\mu\nu} = A(\tilde X)^{-1} \tilde g^{\mu\nu}$ as an ansatz for the inverse metric. Direct substitution in $\tilde g_{\mu\nu}=C(X)g_{\mu\nu}$, $X=-\half g^{\al\bt}\phi_{,\al}\phi_{,\bt}$ yields the following condition on the form of $A(\tilde X)$
\begin{equation}\label{inv_conf1}
C\big( \tilde X/A(\tilde X)\big) A(\tilde X) = 1\,,
\end{equation}
with $X=\tilde X/A(\tilde X)$, or alternatively $A(\tilde X)=1/C(X)$ (by contracting $\phi_{,\mu}\phi_{,\nu}$ with both metrics one obtains the equalities $\tilde X = X/C(X)$ and $X=\tilde X/A(\tilde X)$). Note that $A(\tilde X)$ can be a multi-valued function, as shown in figure \ref{fig_inverseConfExp}.

% %Simple examples
In order to see this formalism at work, let us consider an exponential derivative conformal factor $C=\exp(X/M^4)$. This choice ensures that $C\neq 0$ for any finite $X$, but the additional eigenvalue $\la_K = C(1-X/M^4)$ spoils the existence of an inverse around points where $X= M^4$. 
The relation for the inverse conformal factor, Eq. (\ref{inv_conf1}), reduces to $\tilde X/M^4 = -A(\tilde X)\log(A(\tilde X))$. It is possible to obtain $A(\tilde X)$ implicitly, as shown in the left panel of Figure \ref{fig_inverseConfExp}, where it is clear that $A(\tilde X)$ becomes bi-valuate for $\tilde X/M^4 \in [0,e^{-1})$. The point $\tilde X/M^4 = e^{-1}$ is an upper bound on the field gradient, which corresponds to the singular point $X= M^4$ at which $\la_K=0$. 

As an example of a better behaved relation, one may consider a Gaussian function $C=\exp(-\half X^2/M^8)$. Unlike in the previous case, this choice of the conformal factor ensures that neither of the eigenvalues (\ref{conf_eigen1}, \ref{conf_eigen2}) vanish, as $C>0$ and $C-C_{,X}X = C(1+X^2/M^8)>0$. The inverse relation (\ref{inv_conf1}) satisfies $\tilde X^2/M^4 = A^2(\tilde X)\log(A^2(\tilde X))$, and therefore for any solution $A(\tilde X)$, another solution $-A(\tilde X)$ exists. However, the two branches do not meet, as shown in the right panel of Figure \ref{fig_inverseConfExp}. These examples show how the Jacobian analysis can provide a valuable tool to analyze scalar-tensor theories with general derivative couplings to matter.

\subsection{General Disformal Relation} \label{inverse_D_X}

The arguments can be straightforwardly generalized to the disformal relation (\ref{eq:disf_Bekenstein}). The Jacobian  reads
\begin{equation}\label{eq:jacobian_general}
\frac{\partial \tilde g_{\mu\nu}}{\partial g_{\al\bt}} =
C\delta^\al_\mu\delta^\bt_\nu + \half C_{,X}g_{\mu\nu}\phi^{,\al}\phi^{\bt} 
+ \half D_{,X}\phi_{,\mu}\phi_{,\nu}\phi^{,\al}\phi^{,\bt}\,,
\end{equation}
where we have again omitted the possible dependence on $\phi$, which does not modify the equations. 
The equation for the eigenvalues (\ref{eigenvalues}) reads
\begin{equation}
 (C-\la)\xi_{\mu\nu} + \half \lp C_{,X}g_{\mu\nu} + D_{,X}\phi_{,\mu}\phi_{,\nu} \rp \enangle{\xi_{\al\bt}} =0\,,
\end{equation}
and yields the following set of eigentensors:
\begin{eqnarray}
\la_C=C,\phantom{- C_{,X}X + 2 D_{,X} X^2\, } &\;&  \xi^{C}_{\mu\nu} = v^{(1)}_{(\mu}v^{(2)}_{\nu)}, 
\quad \text{with } v^{(n)}_\mu\phi^{,\mu} =0 \label{disf_eigen1}
\,, \\
\la_K=  C - C_{,X}X + 2 D_{,X} X^2 ,  & & \xi^K_{\mu\nu} =   C_{,X} g_{\mu\nu} + {D_{,X}}\phi_{,\mu}\phi_{,\nu}   \label{disf_eigen2}
\end{eqnarray}
The conformal eigenvalue and its associated eigentensor (\ref{disf_eigen1}) have the same expression as it was found in the previous section for a pure conformal coupling (\ref{conf_eigen1}), while the kinetic eigenvalue and eigentensor (\ref{disf_eigen2}) are modified if $D_{,X}\neq 0$. Just as in the conformal case, $\la_C$ is degenerate with multiplicity 9 and $\la_K$ is non-degenerate unless $X$ or $C_{,X},D_{,X}$ are zero. Note that $\xi^K_{\mu\nu}$ coincides with the partial derivative of $\tilde g_{\mu\nu}$ with respect to $X$.

Any values of $X$ for which $\la_C,\la_K$ become zero indicate the lack of existence of an inverse transformation. The term introduced by the disformal part of the transformation is proportional to $X^2$ rather than linear in $X$. Part of the difficulties in finding a suitable (purely) conformal function in the previous section were that $X$ can have either sign depending on whether $\phi_\mu$ is timelike (+) or spacelike (-). Adding a disformal factor with $D_{,X}>C_{,X}/X-C/X^2$ may prevent the singular points from occurring. In particular, a purely disformal transformation ($C=1$) has viable eigenvalues provided that $D_{,X}$ is positive.
However, disformal relations with $\la_K\neq 0$ might still be problematic if the determinant of the metric (\ref{eq:disf_det}) vanishes, in which case the contravariant metric $\tilde g^{\mu\nu}$ (and therefore $g^{\mu\nu}$) is not well defined.

\begin{figure*}[t]
 \includegraphics[width=0.47\textwidth]{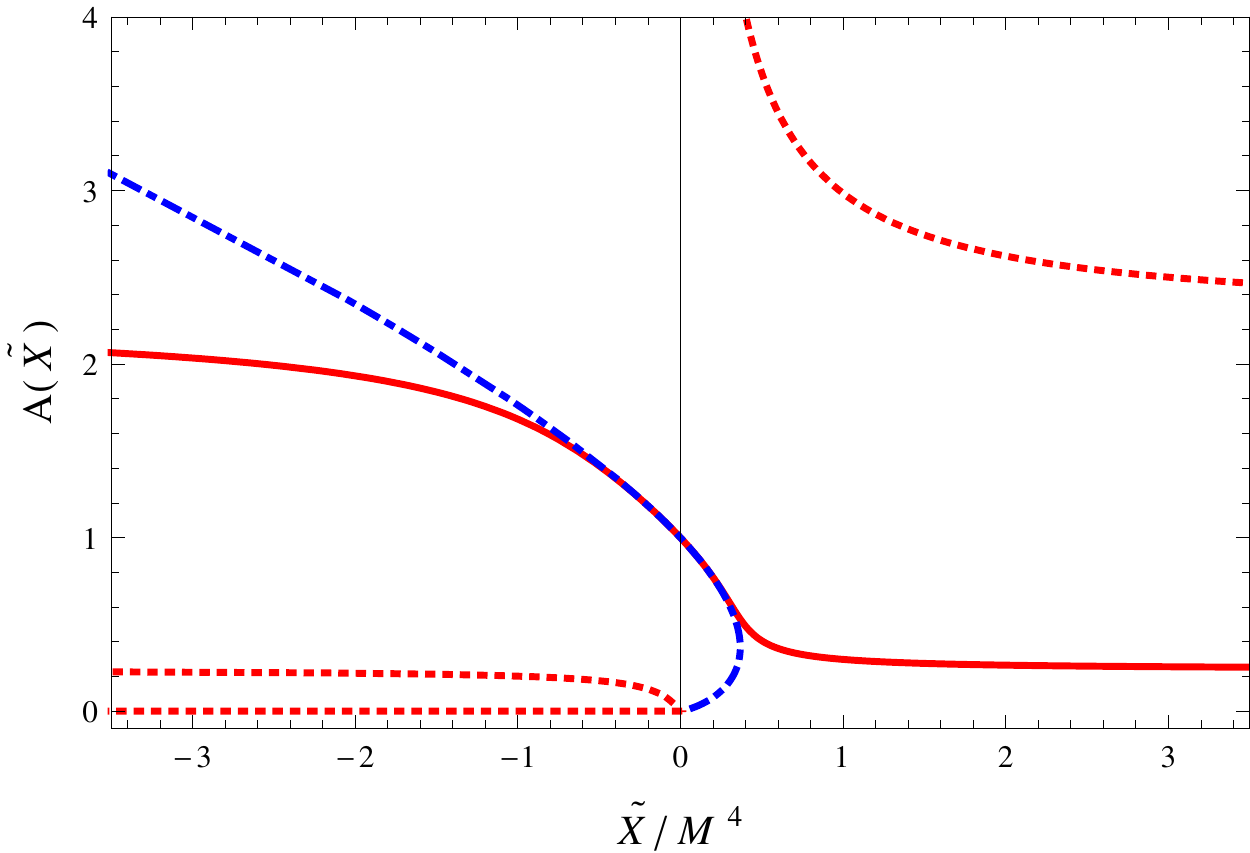} \hfill
 \includegraphics[width=0.47\textwidth]{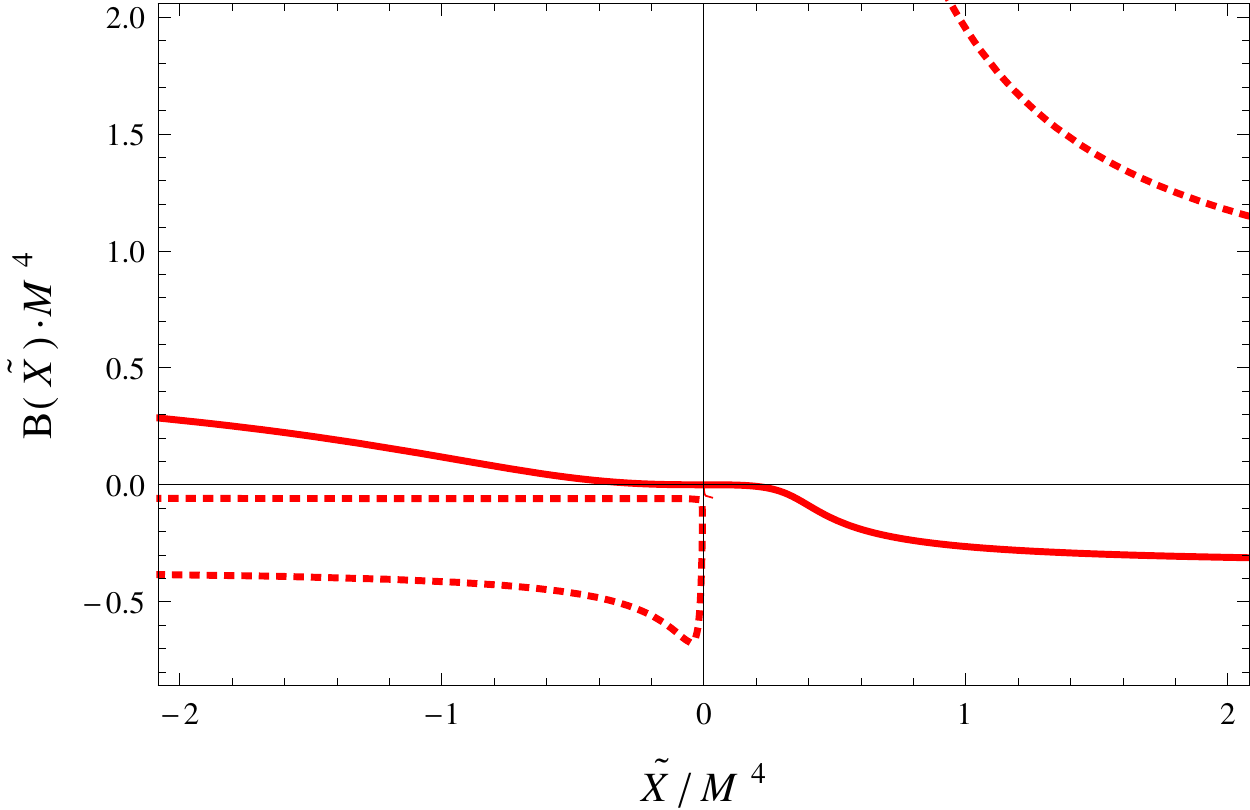}\hspace{2pt}
\caption{Inverse map for $\tilde g_{\mu\nu} = \exp(X/M^4)g_{\mu\nu} + \half X^3/M^{16}\phi_{,\mu}\phi_{,\nu}$ given by Eq. (\ref{inv_disf1}) with the inverse conformal function $A(\tilde X)$ on the left panel and the inverse disformal function $B(\tilde X)$ on the right. Both functions admit four branches with positive (solid) and negative (dotted) values of $C(X)-2D(X)X$, due to the fact that $\tilde X(X)$ has multiple poles. The inverse conformal factor of $\tilde g_{\mu\nu} = \exp(X/M^4)g_{\mu\nu}$ (left panel of figure \ref{fig_inverseConfExp}) is shown for comparison (blue dash-dotted). A branch with $C(X)-2D(X)X>0$ exists for large values of $X$, but is difficult to plot because the inverse conformal factor tends to zero rapidly.}
\label{fig_inverseDisf}
\end{figure*}

An inverse map for the lowercase metric, $g_{\mu\nu}(\tilde g_{\al\bt})$ might be found around any non-degenerate point. It should be of the disformal type to ensure that the causal distortion induced by the transformation is proportional to $\phi_{,\mu}$. Therefore the ansatz for the inverse is $g_{\mu\nu} = A(\tilde X) \tilde g_{\mu\nu} + B(\tilde X) \phi_{,\mu}\phi_{,\nu}$, $g^{\mu\nu} = \frac{1}{A(\tilde X)}\lp \tilde g^{\mu\nu} - \frac{B(\tilde X)}{A(\tilde X)-2B(\tilde X)\tilde X}\phi^{,\mu}\phi^{,\nu}\rp$ and $\tilde X = -\half \tilde g^{\al\bt}\phi_{,\al}\phi_{,\bt}$. Substituting the original metric in this expression yields the conditions
\begin{eqnarray}  \label{inv_disf1}
A(\tilde X)=\frac{1}{C(X)}\,,
\quad B(\tilde X) = -\frac{D(X)}{C(X)}\,, 
\quad \tilde X  = \frac{X}{C-2DX}\,.
\end{eqnarray} 
This relation for the field's kinetic terms obtained with respect to the two metrics becomes singular when the relation between the determinants (\ref{eq:disf_det}) does. Figure \ref{fig_inverseDisf} shows the inverse of a transformation exhibiting this type of singularity, which is related to its multi-valued character. It is still possible that these singularities are dynamically avoided, and the inverse map remains within a certain branch. This has been found for special disformal relations, as it was discussed after Eq. (\ref{eq:disf_det}).

\section{The Jordan Frame: Second Order Theories beyond the Horndeski Lagrangian} \label{sect:frame_Jordan}

%Brief summary
In previous sections it was shown that
1) a scalar field coupled to matter through a disformal metric (\ref{eq:disf_Bekenstein}) is described by second order equations (\ref{eq:Q_Bekenstein}) and 2) an inverse map to the Jordan frame exists, except around points for which (\ref{disf_eigen1}) or (\ref{disf_eigen2}) vanish.
These results seem in contradiction with the finding of non-Horndeski terms introduced by general disformal transformations of the gravitational sector, unless these are of the special disformal type (\ref{eq:disf_special}) \cite{Bettoni:2013cba}.

%Action
The purpose of this section is to examine this apparent contradiction by explicitly computing the Jordan frame action and equations. For the sake of concreteness, the analysis will be restricted to theories in which the original frame, for which the equations of section \ref{section:beyondBekenstein} hold, is actually the Einstein frame. The action is then given by
\begin{equation}\label{eq:EH_action}
 S_J[g_{\mu\nu},\phi,\psi_m] = \int d^4 x \left\{ \sqrt{- \bar g} \frac{ \bar R [ \bar g_{\mu\nu}]}{16\pi G} + \sqrt{- g}\Lag_{M}[ g_{\mu\nu},\psi_m] + \sqrt{-g}\Lag_\phi[g_{\mu\nu},\phi] \right\} \,,
\end{equation}
where  $g_{\mu\nu}$, $\phi$ and the matter fields $\psi_m$ are the dynamical variables which determine the field equations, and the gravitational metric $\bar g_{\mu\nu}[g_{\mu\nu},\phi]$ is given by the inverse mapping of the metric to which matter couples (e.g. Eq. (\ref{eq:disf_Bekenstein})), as discussed in the previous section.
It will be assumed that the scalar field Lagrangian is described by terms of the form (\ref{LH2}, \ref{LH3}), on which a disformal transformation only changes the form of the free functions (cf. (\ref{disfL2}, \ref{disfL3}) for a simpler case).

%Expanding the action
The transformed gravitational action can be written in terms of the difference between the covariant derivatives associated with $\bar g_{\mu\nu}$ and $g_{\mu\nu}$
\begin{eqnarray} \label{connections}
\mathcal{K}^{\al}_{\ph \mu\nu} &\equiv&  \bar\Gamma^{\al}_{\mu\nu} - \Gamma^{\al}_{\mu\nu}
= \bar g^{\al\la} \Big( \nabla_{(\mu}\bar g_{\nu)\la} -\half\nabla_\la \bar g_{\mu\nu} \Big)\,,
\end{eqnarray}
which transforms as a tensor \cite{Wald_book}.
The barred Riemann tensor can then be defined from the commutator of covariant derivatives acting on a vector $ 2 \bar \nabla_{[\mu} \bar \nabla_{\nu]} v^{\al} \equiv \bar R\ud{\al}{\bt\mu\nu} v^\bt $, which allows one to write
\begin{eqnarray}\label{riemmangen}
\bar R^\al_{\phantom{\al}\bt\mu\nu} &=&  
R^\al_{\ph\bt\mu\nu}
 + 2\nabla_{[\mu}\mc K^{\al}_{\ph \nu]\bt}
+ 2\mc K ^\al_{\ph \gamma [\mu} \mc K^\gamma _{\ph \nu]\bt}\,, 
\end{eqnarray}
as well as its contractions, such as the Ricci scalar:
\begin{equation}\label{riccigen}
\bar R \equiv \bar g^{\mu\nu} \bar R^\al_{\ph\mu\al\nu} \,,
\end{equation}
where the barred metric has to be used self-consistently, e.g. $\bar R_{\al\bt\mu\nu}\equiv \bar g_{\al\la}\bar R\ud{\la}{\bt\mu\nu}$, $\bar R^{\al\bt\mu\nu}\equiv \bar g^{\bt\la}\bar g^{\mu\sigma}\bar g^{\nu\kappa} \bar R\ud{\al}{\la\sigma\kappa}$.
%further simplifications
It is possible to express the unbarred covariant derivative in the second term of Eq. (\ref{riemmangen}) as a barred covariant derivative using the relation $(\bar \nabla_\al- \nabla_\al) \mc K\ud{\la}{\mu\nu} = \mc K\ud{\la}{\al\sigma} \mc K\ud{\sigma}{\mu\nu}- 2 \mc K\ud{\sigma}{\al(\mu} \mc K\ud{\la}{\nu)\sigma}$.
As $\bar \nabla_\al\bar g^{\mu\nu}=0$, an equivalent expression for the Jordan frame action (\ref{eq:EH_action}) can be obtained
\begin{eqnarray}\label{eq:EH_Kfinal}
 S_{J}[g_{\mu\nu},\phi,\psi_m] &=& 
\int d^4 x  \left\{\frac{\sqrt{- \bar g}}{16\pi G}\lp \bar g^{\mu\nu}\lp  R\ud{\al}{\mu\al\nu} 
- 2 \mc K ^\al_{\ph \gamma [\al} \mc K^\gamma _{\ph \mu]\nu} \rp 
+ \bar\nabla_{\al}\xi^\al \rp
+ \sqrt{- g}\Lag_{M} + \sqrt{-g}\Lag_\phi \right\} 
\end{eqnarray}
where the dependences on the right hand side have been dropped and $\xi^{\al} \equiv  \mc K\ud{\al}{\mu\nu} \bar g^{\mu\nu} - \bar g^{\al\mu} \mc K\ud{\nu}{\mu\nu} $ enters through a total derivative and thus does not contribute to the equations of motion.
This expression has the advantage of not introducing derivatives of the connection, and therefore keeps only second derivatives of the metric and the scalar field in the Lagrangian. Let us consider the case of a derivative conformal relation before tackling the general case.

\subsection{Derivative Conformal Relation}\label{section:JordanCX}

The simplest case that can be considered is the gravitational Lagrangian of a pure conformal relation 
\begin{equation}
 \bar g_{\mu\nu} = \Omega^2(X,\phi) g_{\mu\nu}\,,
\end{equation}
(the conformal factor is squared in order to simplify the equations and facilitate the comparison with the literature).
The action in the Jordan Frame is given by $S_C[g_{\mu\nu},\phi,\psi_m]=\int d^4 x \Lag_{C}$, with
\begin{equation}\label{LagConformal}
\Lag_{C} = \frac{\sqrt{-g}}{16\pi G} \lp \Omega^2 R + 6 \Omega_{,\al}\Omega^{,\al} \rp  + \sqrt{-g}\lp \Lag_\phi + \Lag_m \rp\,,
\end{equation}
after integration by parts as in Eq. (\ref{eq:EH_Kfinal}). As it was noted in Ref. \cite{Bettoni:2013cba}, the second term contains $6 (\Omega_{,X})^2 \enangle{\Phi}$, which can not be written in the appropriate Horndeski form (\ref{LH4}). As a consequence, its variation with respect to the field contains up to fourth field derivatives: $\frac{\delta \Lag_C}{\delta \phi} \hashd \nabla_{\mu}\nabla_{\nu} \frac{\partial \Lag_C}{\partial \phi_{;\mu\nu}} \hashd (\Omega_{,X})^2 \phi_{,\sigma}\phi\ud{;\sigma\mu}{;\mu\nu}\phi^{,\nu}$, where $\hashd$ denotes that only the higher derivatives are shown in the right hand side.

Let us examine the full equations of motion in detail. Variation of the Jordan frame Lagrangian (\ref{LagConformal}) yields
\begin{equation}\label{confMetricEq}
 \Omega^2 G_{\mu\nu} + 2\Omega ( g_{\mu\nu}\Box \Omega - \Omega_{;\mu\nu})
+ (6\Box \Omega- \Omega R) \Omega_{,X}\phi_{,\mu}\phi_{,\nu} 
- g_{\mu\nu}\Omega_{,\al}\Omega^{,\al}
  + 4 \Omega_{,\mu}\Omega_{,\nu}
 = 8\pi G (T_{\mu\nu}^\phi + T_{\mu\nu}^m) \,,
\end{equation}
\begin{equation}\label{confKGeq}
 \nabla_\mu\lp \Omega_{,X}\phi^{,\mu}(\Omega R - 6\Box \Omega)\rp + \Omega_{,\phi}(\Omega R - 6\Box \Omega) + \half \frac{\delta \Lag_\phi }{\delta\phi }= 0\,,
\end{equation}
where it has been used that 
$\delta \Omega = \Omega_{,\phi}\delta\phi -\Omega_{,X}(\phi^{,\al}\nabla_\al\delta\phi + \half \phi_{,\al}\phi_{,\bt}\delta g^{\al\bt})$, 
$\delta \sqrt{-g}=-\half \sqrt{-g}g_{\mu\nu}\delta g^{\mu\nu}$ 
and $\delta R = R_{\mu\nu}\delta g^{\mu\nu} + g_{\mu\nu} \Box \delta g^{\mu\nu} - \nabla_\mu\nabla_\nu \delta g^{\mu\nu}$.
Although the field equation (\ref{confKGeq}) does indeed contain the expected fourth order term, a relation between $\Box \Omega$ and $R$ can be obtained by taking the trace of the gravitational equation (\ref{confMetricEq}):
\begin{equation}\label{confGtrace}
 ( 6\Box \Omega - \Omega R )(\Omega-2\Omega_{,X}X) = 8\pi G T\,,
\end{equation}
with $T=g^{\mu\nu}(T_{\mu\nu}^\phi + T_{\mu\nu}^m)$. This relation motivates the definition of the \emph{kinetic mixing factor} for a conformal transformation
\begin{equation}\label{confAcal}
 \mc  T_{\rm K} \equiv\frac{8\pi G  \Omega_{,X} T }{\Omega-2\Omega_{,X}X}\,.
\end{equation}
Using this definition, equation (\ref{confGtrace}) can be substituted back into the original equations, yielding a rather simple result
\begin{equation}\label{confMetricEq2}
 \Omega^2 G_{\mu\nu} +  2 \Omega \lp g_{\mu\nu}\Box \Omega - \Omega_{;\mu\nu}\rp 
+  \mc  T_{\rm K} \phi_{,\mu}\phi_{,\nu} 
-g_{\mu\nu}\Omega_{,\al}\Omega^{,\al} + 4 \Omega_{,\mu}\Omega_{,\nu} = 8\pi G T_{\mu\nu}^{\rm tot}\,,
\end{equation}
\begin{equation}\label{confKGeq2}
 \nabla_\mu\lp \phi^{,\mu} \mc  T_{\rm K} \rp 
+ \frac{\Omega_{,\phi}}{\Omega_{,X}} \mc  T_{\rm K} 
- \half \frac{\delta \Lag_\phi }{\delta\phi }= 0\,,
\end{equation}

The equations do not contain higher than second derivatives of the dynamical variables when written in this form.
The field equation (\ref{confKGeq2}) is manifestly second order, since the term in parenthesis contains at most first derivatives of $\phi$ (as long as $\nabla_\mu( T^\phi + T^m)$ and $\frac{\delta \Lag_\phi }{\delta\phi }$ are themselves second order). Third order time derivatives in the gravitational equations might arise from the second term in (\ref{confMetricEq2}). To show that this is not the case, we can examine its temporal, mixed and spatial components
\begin{equation}\label{eq:highDsreduced}
g_{\mu\nu}\Box \Omega - \Omega_{;\mu\nu} =
\left( \begin{array}{c | c}
g^{k\al}\Omega_{;k\al} & - \Omega_{;0i} \\[1pt]  \hline
& \\ 
 - \Omega_{;0i}  &  \quad g_{ij}\Box \Omega - \Omega_{;ij}
\; \\ &
\end{array} \right)\,,
\end{equation}
where sums over $\alpha=0-3$, $k=1-3$ are implicit. As $\Omega_{;\mu\nu} \hashd \Omega_{,X}\phi^{,\al}\phi_{;\al\mu\nu}$, the only third time derivatives of the field would occur in the spatial part of the tensor through $\Box \Omega$, which is second order by virtue of Eq. (\ref{confGtrace}). The cancellation of the high time derivatives in the $(0,\mu)$ components is to be expected, as they represent constraint equations for $\bar G_{\mu\nu}$ \cite{Wald_book}.

\subsection{General Disformal Relation}\label{section:JordanGenX}

Let us now study the disformal case in which the Jordan frame action is given by Eq. (\ref{eq:EH_Kfinal}) with
\begin{equation}\label{eq:disf_inverse}
\bar g_{\mu\nu}=A(X,\phi) g_{\mu\nu} + B(X,\phi)\phi_{,\mu}\phi_{,\nu}\,,
\end{equation}
the inverse map of (\ref{eq:disf_Bekenstein}), as discussed in section \ref{section:jacobian}.
For transformations beyond the derivative conformal and special disformal cases, the action expressed in terms of the dynamical variables $g_{\mu\nu},\phi$ acquires many terms and the equations become very difficult to handle in practice. However, it is still possible to address the second order nature of the evolution in relatively simple terms if both barred and unbarred quantities are allowed in the equations.%
\footnote{This is similar to the analysis of Ref. \cite{RenauxPetel:2011uk} for inflationary scenarios.}
The Jordan frame form of the resulting theory will be shown in Appendix \ref{beyondHorndeski}.

The variation of the gravitational sector (\ref{eq:EH_action}) can be expressed in terms of barred quantities, in which it has the usual General Relativistic form
\begin{equation}\label{eq:bar_grav_var0}
 \delta \lp \sqrt{-\bar g} \bar R \rp = 
-\sqrt{-\bar g}\lp \bar G^{\mu\nu}\delta \bar g_{\mu\nu} + (\bar g^{\mu\nu}\bar\square - \bar\nabla_\mu\bar\nabla_\nu)\delta \bar g_{\mu\nu}\rp\,,
\end{equation}
The second term is a total derivative and hence does not contribute to the equations of motion.%
\footnote{The structure of the variation (\ref{eq:bar_grav_var0}) strongly suggests that the appropriate boundary term necessary to ensure that $\nabla_\alpha \delta g_{\mu\nu}=0$ on the boundary \cite{York:1972sj,Gibbons:1976ue}, would be reproduced from the original one when transforming to the Jordan frame. See Ref. \cite{Saltas:2010ga} for the study of this term in $f(R)$ and Gauss-Bonnet gravity.}
We can now write the equations in terms of the dynamical variables. 
The variation of the barred metric reads
\begin{equation}\label{eq:bar_metric_var}
\delta \bar g_{\mu\nu}=\frac{\partial \bar g_{\al\bt}}{\partial g_{\mu\nu}}\delta g_{\al\bt} 
 - \lp A_{,X} g_{\mu\nu} + B_{,X}\phi_{,\mu}\phi_{,\nu} \rp \phi^{,\al}(\delta\phi)_{,\al}
+ 2 B\phi_{,(\mu}(\delta\phi)_{,\nu)}
+  \lp A_{,\phi} g_{\mu\nu} + B_{,\phi}\phi_{,\nu}\phi_{,\mu}\rp\delta\phi\,.
\end{equation}
Note that the metric part of the variation is just given by the Jacobian (\ref{eq:jacobian_general}) and the first term in parenthesis is proportional to the kinetic eigentensor $\xi^K_{\mu\nu}$, given by Eq. (\ref{disf_eigen2}). The last terms are the ones arising from a special disformal relation. The connection between the variation and the Jacobian has strong consequences for the structure of the equations.

The equations for gravity and the scalar field can then be written using (\ref{eq:bar_grav_var0}, \ref{eq:bar_metric_var}) in terms of the Jacobian (\ref{eq:jacobian_general}) and the kinetic eigentensor $\xi^K_{\mu\nu}$ (\ref{disf_eigen2})
\begin{equation}\label{eq:grav_Jordan_gen}
\bar G^{\al\bt}\frac{\partial\bar g_{\al\bt}}{\partial g_{\mu\nu}} = 8\pi G\sqrt{\frac{g}{\bar g}} \lp T^{\mu\nu}_m + T^{\mu\nu}_\phi\rp \,,
\end{equation}
\begin{equation}\label{eq:field_Jordan_gen}
 \bar \nabla_\al\lp \bar G^{\mu\nu} \xi^K_{\mu\nu}\phi^{,\al}\rp - \bar G^{\mu\nu}\bar \nabla_\mu\lp B\phi_{,\nu}\rp 
+ \bar G^{\mu\nu}\lp A_{,\phi} g_{\mu\nu} + B_{,\phi}\phi_{,\nu}\phi_{,\mu}\rp 
- \sqrt{\frac{g}{\bar g}}\frac{\delta \Lag_\phi}{\delta \phi} = 0\,,
\end{equation}
where in the field equation the barred Bianchi identity $\bar \nabla_\mu \bar G^{\mu\nu}=0$ has been used on the second term.
The analogue of Eq. (\ref{confGtrace}) which allows to solve for the higher derivatives can be obtained by contracting (\ref{eq:grav_Jordan_gen}) with the kinetic eigentensor
\begin{equation}\label{eq:kinmix_JF_gen}
 \bar G^{\mu\nu} \xi^K_{\mu\nu} =  8\pi G \sqrt{\frac{g}{\bar g}}\frac{T^{\mu\nu}_{\rm tot}\, \xi^K_{\mu\nu}}{\la_K}
\equiv \mc T_{\rm K}
\end{equation}
where the kinetic eigenvalue $\la_K$ comes from the action of the Jacobian on its eigentensor $\xi^K_{\mu\nu}$.
This equation provides the generalization of the kinetic mixing factor obtained for the purely conformal case (\ref{confAcal}), which played a central role in the reduction of the equations to a second order expression in the previous section. 
Divided by $G$, it provides a dimensionless measure of the kinetic mixing between the scalar field and matter due to the $X$ dependence of the barred metric: it vanishes identically both in vacuum or for theories with $A_{,X},B_{,X}=0$. In addition, it diverges at points in which either the kinetic eigenvalue or the barred metric become singular. Note that it is still possible to have kinetic mixing and $\mc T_{\rm K}=0$ if the metric is of the special disformal type (\ref{eq:disf_special}).

%Remove high time derivatives
Equation (\ref{eq:kinmix_JF_gen}) allows one to write the field equation (\ref{eq:field_Jordan_gen}) in a manifestly second order form
\begin{equation}\label{eq:field_Jordan_gen2}
\bar \nabla_\al\lp \mc T_{\rm K} \phi^{,\al}\rp - \bar G^{\mu\nu}\bar \nabla_\mu\lp B\phi_{,\nu}\rp 
+ \bar G^{\mu\nu}\lp A_{,\phi} g_{\mu\nu} + B_{,\phi}\phi_{,\nu}\phi_{,\mu}\rp 
- \sqrt{\frac{g}{\bar g}}\frac{\delta \Lag_\phi}{\delta \phi} = 0\,,
\end{equation}
since $\mc T_{\rm K}$ only contains first derivatives of the field.
It remains to show that the third time derivatives present in $\bar G^{\mu\nu}$ can be solved away using (\ref{eq:kinmix_JF_gen}).
The higher order structure of the barred Ricci tensor is given by the $2\nabla_{[\la}\mc K\ud{\la}{\bt]\al}$ terms from (\ref{riemmangen}), which can be expanded
\begin{eqnarray}\label{eq:Rhighder0}
\bar R_{\al\bt} &\hashd& 
\half  \bar g^{\la\sigma} \lp \bar g_{\sigma\bt,\al\la} + \bar g_{\sigma\al,\bt\la} -\bar g_{\al\bt,\la\sigma} - \bar g_{\la\sigma,\al\bt}\rp 
 \,\hashd\,
 \half  \bar g^{\la\sigma} 
\lp \xi^K_{\sigma\bt}X_{,\al\la} +  \xi^K_{\sigma\al}X_{,\bt\la} -\xi^K_{\al\bt}X_{,\la\sigma} - \xi^K_{\la\sigma}X_{,\al\bt}\rp\,.
\label{eq:Rhighder}
\end{eqnarray}
The second relation follows from introducing $\bar g_{\bt\sigma,\al\la} \hashd  2 B\phi_{,(\sigma}\phi_{,\bt)\al\la} + \xi^K_{\sigma\bt}X_{,\al\la}$ in the first term in Eq. (\ref{eq:Rhighder}), with the contribution proportional to $B$ cancels due to antisymmetry in $\bt,\la$.%Special disformal relations 
\footnote{The reason why special disformal transformations do not introduce non-Horndeski terms in the action \cite{Bettoni:2013cba}, while $X$-dependent disformal maps do, is due to the fact that higher field derivatives in the barred Ricci tensor are proportional to $\xi_{\mu\nu}^{K}$. In the special case the equations of motion only contain up to second field derivatives without the need to use an implicit relation such as Eq. (\ref{eq:kinmix_JF_gen}), as they depend on $\bar R_{\mu\nu}$. Theories for which $\xi_{\mu\nu}^K\neq 0$, they must belong to the set of theories described by Horndeski's theorem.}
The higher derivative structure of the barred Einstein tensor is given by
\begin{eqnarray}\label{eq:einsteinhd}
\bar G^{\al\bt} \hashd \bar G^{\al\bt}_{(!)} & \equiv & 
\half \left\{2\bar \xi^{\delta(\al}\bar g^{\bt)\ga} - \bar \xi\, \bar g^{\al\ga}\bar g^{\bt\delta} - \bar \xi^{\al\bt}\bar g^{\ga\delta}
- \bar g^{\al\bt}\lp \bar \xi^{\ga\delta}- \bar \xi\, \bar g^{\ga\delta} \rp
 \right\}{X_{,\ga\delta}}\,,
\end{eqnarray}
where $\bar \xi^{\al\bt} = \bar g^{\al\mu}\bar g^{\bt\nu}\xi^K_{\mu\nu}$ and $\bar \xi = \bar g^{\mu\nu}\xi^K_{\mu\nu}$. Third time derivatives in the above expression occur only through $X_{,00}=\phi^{,\sigma}\phi_{,\sigma 00}$. The value of $X_{,00}$ can be solved for in terms of lower derivatives using the constraint equation (\ref{eq:kinmix_JF_gen}) together with (\ref{eq:einsteinhd}) (problems associated to points at which the coefficient of $X_{,00}$ approaches zero might occur, analogous to the Jacobian eigenvalues becoming zero).
%initial value problem
Although the third time derivatives of the field can be generically removed from the equations in theories with $\xi_{\mu\nu}^K\neq 0$, Eq. (\ref{eq:einsteinhd}) still contains third field derivatives involving spatial directions. This feature might be relevant for the initial value problem in this type of non-Horndeski theories, although such derivatives are absent in the Einstein frame equations, cf. Section \ref{section:beyondBekenstein}. The initial value formulation in disformally coupled theories deserves a detailed study, which will be presented elsewhere.

\subsection{A Loophole in Horndeski's Theorem: Hidden Constraints}

%Implications: The loophole is given by Implicit constraints
The computations in the last subsections show how the higher time derivatives of the scalar field can be eliminated from the Jordan frame equations through the use of implicit constraint relations, rendering the system second order. A direct implication is the incompleteness of Horndeski's theorem in the identification of Ostrogradski stable scalar-tensor theories, since it only identifies the maximal set of theories with a second order Euler-Lagrange equations (\ref{eq:euler_lagrange}), regardless of implicit constraint relations that allow to solve for the higher derivatives.
The possibility of re-writing apparently higher order theories in a second order form is not explored in the different proofs of Horndeski's theorem \cite{Horndeski,Deffayet:2009mn,Deffayet:2011gz}.
Alternatively, the loophole can be regarded as the possibility of allowing the scalar field to directly modify the matter metric, e.g. allowing a general disformal coupling to matter (\ref{eq:disf_Bekenstein}).

%Phase space, and exorcising Ostrogradski's ghost
The use of implicit (constraint) relations to remove the higher time derivatives of the dynamical variables is related to a recent result found in \cite{Chen:2012au}, which allows to ``exorcise Ostrogradski's ghost in higher order derivatives with constraints''. It states that theories with constraints which reduce the dimensionality of the phase space of the system do not have a linear instability in the Hamiltonian, even if the original Lagrangian includes second or higher derivatives with respect to time. This result is obtained in the context of one dimensional, higher order theories with constraints in the form of Lagrangian multipliers. The extension to four dimensional field theories with implicit constraints seems plausible, as the constraints determine the value of the field's third and fourth derivatives, effectively reducing the dimension of the phase space by two. This strongly suggests that Ostrogradski's theorem does not apply to the Jordan frame representation of disformally coupled theories, 
in agreement with the second order description in the Einstein frame shown in section \ref{einsteinframe}. 

%f(R) as an example of basically the same
Another well known example of na\"ivelly unstable field theories is given by $f(R)$ gravity, to which Ostrogradski's theorem would in principle apply \cite{Woodard:2006nt}. The Euler-Lagrange equations for such theories yield up to quartic derivatives of the metric through the terms $\nabla_\mu\nabla_\nu f_{,R}- g_{\mu\nu}\square f_{,R}$. As higher derivatives of the metric always occur in terms of first or second derivatives of $f_{,R}$, the loophole to Ostrogradski's theorem comes from the identification of $f_{,R}$ as a new scalar degree of freedom. However, this propagating field is not associated to an unbounded Hamiltonian. Then, although $f(R)$ gravity does not formally belong to Horndeski's theory, it can be shown to be equivalent to a scalar-tensor theory specified by $G_{4,X}=G_3=G_5=0$, $G_2\neq 0$ by means of a Legendre transformation \cite{Barrow:1988xh,Maeda:1988ab}.

%Veiled General Relativity
Another example of an implicit constraint occurs in ``veiled'' General Relativity, in which both the matter and the gravitational sector are conformally transformed by $g_{\mu\nu}\to A(\phi) g_{\mu\nu}$, see \cite{Deruelle:2010ht}. In the transformed frame, the equation for the conformal factor $A(\phi)$ reduces to the trace of the metric equations. This shows that the dynamics of the scalar field are redundant, as expected from the fact that the scalar field was introduced artificially (and not present in the original frame). The case of ``veiled'' GR is analogous (although simpler) to the Jordan frame version of disformally coupled theories: taking the trace of the metric equations is equivalent to the contraction with the kinetic eigentensor discussed in sections \ref{section:JordanCX} and \ref{section:JordanGenX}. In disformally coupled theories only the higher order dynamics are artificial, and can consequently be eliminated by a similar procedure.

%Other cases
A hint about the incompleteness of Horndeski's theory has been obtained in the framework of effective field theory for (linear) cosmological perturbations, in the form of an operator combination which leads to second order equations but can not be obtained from any of the terms in $\Lag_H$ (\ref{LH2}-\ref{LH5}) \cite{Gleyzes:2013ooa}. Certain generalized Galileon and massive gravity theories in the decoupling limit also contain a scalar field with a higher derivative Lagrangian, which nonetheless does not introduce additional or ghostly degrees of freedom \cite{Gabadadze:2012tr}. This is shown by writing down the Hamiltonian and finding the primary constraints, leading to a theory with higher \emph{spatial} derivatives in the action. This is analogous to what is found in the theories here studied after applying the implicit constraints, cf. (\ref{eq:highDsreduced}).

%Generality: from EH to Horndeski
The second order nature of the equations has been established under the assumption of an Einstein-Hilbert form for the metric in the original frame. A natural question is whether this assumption is relevant to the procedure, i.e. if the reduction to second order can be applied to the transformed version of more general Lagrangians. This seems plausible in the case of Horndeski's theory, as the gravitational equations do not contain derivatives of the curvature tensor. Therefore, the variation with respect to the dynamical metric is proportional to the Jacobian, as in Eq. (\ref{eq:bar_grav_var0}). The contraction with the kinetic eigentensor will then be proportional to the kinetic eigenvalue, providing an analogue of Eq. (\ref{eq:kinmix_JF_gen}) to substitute the higher derivatives in the field equation with second order terms from the metric equations.

\section{Discussion} \label{section:discuss}

In this work we have examined scalar-tensor theories with derivative couplings to matter, which enter the Lagrangian through an effective metric which depends on the scalar field. Apart from providing a generalization of Jordan-Brans-Dicke theories, they also offer interesting phenomenological possibilities, such as allowing for complete or softly broken shift symmetry and mixing derivatives of the scalar and matter degrees of freedom in the dynamical equations. At a more fundamental level, redefinitions of the physical variables can establish equivalences between classical theories, which can be used to simplify the analysis and provide additional understanding of underlying structures.

The possible relations between the matter and the gravitational metrics can be classified by their tensor structure into conformal (a scalar times the gravitational metric), disformal (the tensor product of a vector) and extended disformal (a rank two tensor whose contraction with the former terms is zero). A complementary classification is provided by the order of the scalar field derivatives introduced in the matter metric. Allowing the relation to depend on second (covariant) derivatives of the scalar introduces a number of difficulties: an (a priori) infinite number of tensor structures can be included, due to the possibility of contracting the field with itself. Covariance requires the introduction of derivatives of the metric through the connection coefficients, which invalidates the algebraic treatment performed here. Finally, such a metric coupling will generically lead to higher derivatives in the equations of motion. This set of arguments single out the disformal relation originally 
introduced by Bekenstein (\ref{eq:disf_Bekenstein}) as the most reasonable choice.

Further physical insight on these theories can be gained by expressing the action in the Jordan frame, i.e. using the metric to which matter couples minimally as a dynamical variable. The mapping to the Jordan frame amounts to inverting the disformal relation, whose existence can be determined studying the Jacobian of the transformation. The inverse map fails to exist around points at which its determinant vanishes, i.e. when one eigenvalue of the Jacobian is equal to zero. This happens when the conformal factor vanishes, but in the case of general disformal transformations can also occur under different circumstances due to the additional dependence of the free functions on the metric, characterized by the kinetic eigenvalue (\ref{disf_eigen2}). The simplicity of the Jacobian analysis makes it a natural starting point in the study of concrete models, as was shown for several examples.

%Jacobian
The Jacobian of the mapping between frames also appears in the study of different aspects of the theory. It determines the relation between the energy-momentum tensor that represents the matter energy density and momentum fluxes seen by observers (obtained by variation with respect to the matter metric) and the source of the gravitational equations (obtained by variation with respect to the gravitational metric). The Jacobian and its kinetic eigentensor also appears in the Jordan frame equations for the metric and the scalar field. Finally, we expect the Jacobian to play a role at the quantum mechanical level by producing extra surface terms due to the transformation rules of the path integral, in a manner analogous to the occurrence of quantum anomalies \cite{Fujikawa:1980vr}. The analysis of this feature might shed some light on the problem of the classical equivalence and quantum inequivalence of physical frames, and is left for a future publication.

%reduction to second order
Disformally coupled theories expressed in the Jordan frame produce terms that do not pertain to the Horndeski Lagrangian, and hence their Euler-Lagrange variation introduces higher derivatives in the equations of motion (unlike in the original frame). However, it is possible to obtain a relation for the higher derivatives by contracting the metric equations with the kinetic eigentensor of the Jacobian. This implicit constraint can be then used to rewrite the dynamics in terms of second order equations, without higher derivatives with respect to time and hence free of Ostrogradski instabilities. The case of a derivative dependent conformal transformation is particularly simple to analyze, as the higher derivatives can be eliminated by taking the trace of the metric equations. 
%Special disformal relations are special
The study of the general case makes clear why special disformal transformations avoid all these difficulties and incarnate a formal invariance of Horndeski's theory \cite{Bettoni:2013cba}: if the free functions only depend on $\phi$, the Jordan frame equations (\ref{eq:field_Jordan_gen}) remain second order as a consequence of the Bianchi identities for the field dependent metric, while in the Einstein frame the equations (\ref{eq:Q_Bekenstein}) simplify due to stress energy conservation with respect to the field dependent metric. This is analogous to the much simpler structure of $\Lag_4,\,\Lag_5$ in Horndeski's theory (\ref{LH4}, \ref{LH5}) when $G_4,\,G_5$ are functions of $\phi$ only.

%The loophole: Sanity of these theories
The analysis of the equations uncovers a loophole in Horndeski's theorem: certain theories, whose variation contains higher derivatives of the fields, might be rendered second order by the existence of hidden constraints in the dynamical equations. Such theories provide further ways to overcome the difficulties generically caused by higher derivative Lagrangians, including the existence of Ostrogradski's instability. This situation shares essential analogies with $f(R)$ gravity (which can be reduced to a second order form by identifying $f_{,R}$ as a scalar degree of freedom) and General Relativity expressed in a different conformal frame (which introduces redundant equations). The reduction of the fields phase space due to constraints has been explicitly shown to eliminate Ostrogradki's ghosts in one dimensional systems \cite{Chen:2012au}, strongly suggesting that this will also be the case for the scalar-tensor theories under consideration. The sanity of disformally coupled theories is further supported by 
the second order nature of the equations in the Einstein frame.

%Looking further
The most immediate question is whether disformally coupled theories represent the most general set of second order theories beyond the Horndeski Lagrangian.%
\footnote{Other ways to generalize Horndeski's theory without introducing ghosts include multi-scalar field theories \cite{Kobayashi:2013ina,Padilla:2012dx,Sivanesan:2013tba}, non-local gravity theories \cite{Biswas:2011ar} and non-linear extensions of Horndeski functions \cite{Appleby:2012rx}}
Classical equivalence between frames implies that any theory which is second order in a given frame will remain second order under field redefinitions. An extension of Horndeski's theorem might be then found in the context of extended disformal transformations which depend on second field derivatives, \emph{if} the free functions displayed in table \ref{table_beyondBekenstein} can be suitably tuned to produce second order equations (although such finely tuned coefficients would be unnatural in a quantum mechanical description if they are not protected by a symmetry). 
The methods presented here can also be applied to study frame transformations in other alternative theories of gravity, including vector-coupled theories such as TeVeS \cite{Bekenstein:2004ne}, conformal vector screening \cite{Jimenez:2012ph} and other generalizations, e.g. \cite{Babichev:2011kq}.
It is also possible that further scalar-tensor theories with hidden constraints can be obtained by modifications in the gravitational sector which can not be absorbed by a redefinition of the metric. A first step in this direction is the study of transformed Gauss-Bonnet term, which is presented in appendix \ref{app:gauss_bonnet}, where it is shown that such terms belong to $\Lag_H$ for special conformal transformations, but not for special disformal transformations.
These and other possibilities (e.g. non-polynomial dependence on second field derivatives) might provide an even larger class of sensible scalar-tensor theories beyond the Horndeski Lagrangian.

%Viability, distinctive character
Theories with implicit constraints therefore constitute a new class of scalar-tensor theories, essentially different from those for which the Euler-Lagrange variation is directly second order, such as Horndeski's theory. The fact that the gravitational equations involve third derivatives of the field (although not third time derivatives) from the barred Einstein tensor might be relevant for the initial value formulation of such theories, even though such a difficulty seems to be absent in the Einstein frame. Lorentz invariance plays a crucial role, for it forces the field derivatives and the metric to occur together and eventually provide the right implicit constraints. More importantly, the dynamical character of spacetime is essential for  the existence of implicit constraints, which are lost if a flat background is imposed. This is in stark contrast with Horndeski theories, for which the Minkowski limit is described by second order equations.
In a broader scope, degenerate field theories might provide new theoretical challenges and phenomenological applications in gravitation and cosmology, as time and time again the search for loopholes in no-go theorems has proved to be a very constructive way to expand the horizons in fundamental physics.

\vspace{10pt}

\textbf{Note added by the authors:} A few weeks before the first version of this manuscript was released, a preprint appeared which had some overlap with some of our results \cite{Bettoni:2013cba}. In particular, their authors present the transformation rules for the Horndenski Lagrangian for special disformal transformations (with arbitrary functions of the field) and show that disformal transformations which depend on $X$ produce non-Horndeski terms. Our computation of the action of special disformal transformations on Horndeski's theory is presented in appendix (\ref{app:special_disf}).

\acknowledgements

We are very thankful to Luca Amendola, Diego Blas, Margarita Garcia P\'erez, Antonio González-Arroyo, Tomi S. Koivisto and Slava Mukhanov for discussions at different stages of this work, to Jose Beltr\'an Jimenez for a very detailed reading of the first manuscript and to Dario Bettoni, Tsutomu Kobayashi, Jeremy Sakstein, Ignacy Sawicki, Filippo Vernizzi and Felix del Teso for correspondence and comments to the first version.
The authors also acknowledge financial support from the Madrid Regional Government (CAM) under the program HEPHACOS S2009/ESP-1473-02, from the Spanish MICINN under grant AYA2009-13936-C06-06 and Consolider-Ingenio 2010 PAU (CSD2007-00060), from the MINECO, Centro de Excelencia Severo Ochoa Programme, under grant SEV-2012-0249, as well as from the European Union Marie Curie Initial Training Network UNILHC PITN-GA-2009-237920. The computations presented in the Appendices have been checked with the xAct package for Mathematica \cite{Brizuela:2008ra,xAct}.

\appendix

\section{General Disformal Relations} \label{app:general_disf}

In this section we will present some relations for disformal relations of the type proposed by Bekenstein
\begin{equation}
 \bar g_{\mu\nu}=Ag_{\mu\nu} + B\phi_{,\mu}\phi_{,\nu}\,,\quad \bar g^{\mu\nu} = \frac{1}{A}\lp g^{\mu\nu} -\ga^2 B\phi^{,\mu}\phi^{,\nu}\rp\,,
\end{equation}
with $\ga^2=(A-2BX)^{-1}$, $X=\half \phi_{,\mu}\phi^{,\mu}$.

\subsection{Connection} \label{App:connection}

The connection for a field dependent metric can be computed directly from the usual definition
\begin{eqnarray}\label{connectionlong}
\bar{\Gamma}^\al_{\mu\nu} 
 &=& \Ga^\al_{\mu\nu} + \delta^\al_{(\mu}{\log{A}}_{,\nu)}-\frac{1}{2}{\log{A}}^{,\al}g_{\mu\nu} 
- \frac{B \ga^2}{A}\phi^{,\al}\Big[ A_{,(\mu}\phi_{,\nu)} - \frac{1}{2}\phi^{,\la} A_{,\la}g_{\mu\nu} \Big]
 \nonumber \\ && \phantom{\Ga^\al_{\mu\nu}}
 + B\ga^2\phi^{,\al}\phi_{;\mu\nu} + \ga^2\phi^{,\al}B_{,(\mu}\phi_{,\nu)} 
 - \frac{1}{2A}\phi_{,\mu}\phi_{,\nu}(B^{,\al}-B\ga^2\phi^{,\al}\phi^{,\la}B_{,\la})
\end{eqnarray}
% \begin{eqnarray}\label{connectionlong}
% \bar{\Gamma}^\mu_{\al\bt} 
% & = & \Ga^\mu_{\alpha\beta} + \delta^\mu_{(\al}{\log{A}}_{,\bt)}-\frac{1}{2}{\log{A}}^{,\mu}g_{\al\bt} 
%  +  \frac{1}{A}\lp \phi^{,\mu} B_{,(\alpha}\phi_{,\bt)} - \frac{1}{2}B^{,\mu}\phi_{,\al} \phi_{,\bt} \rp 
%  \nonumber \\
% && - \frac{B \ga^2}{A}\phi^{,\mu}\Big[ A_{,(\al}\phi_{,\bt)} - \frac{1}{2}\phi^{,\la} A_{,\la}g_{\al\bt} 
%  - 2X\lp B_{,\al}\phi_{,\bt}-\frac{1}{2}\phi^{,\la} B_{,\la}\phi_{,\al} \phi_{,\bt} \rp \Big] \nonumber \\
% && + \frac{B}{A}\Big[ \nabla_{(\al}\lp \phi_{,\bt)}\phi^{,\mu}\rp - \frac{1}{2}\nabla^\mu\lp \phi_{,\al} \phi_{,\bt}\rp - B \ga^2 \phi^{,\mu} \phi^{,\la} \lp \nabla_{(\al}\lp \phi_{,\bt)}\phi_{,\la}\rp - \frac{1}{2}\nabla_\la\lp \phi_{,\al}\phi_{,\bt} \rp \rp\Big]\,, 
% \\ \ga^2 &\equiv& \frac{1}{A-2BX}\,,
% \end{eqnarray}
where $A,B$ are general scalar functions. %Note the unbarred connection coefficients and the partial derivatives combine to form covariant derivatives.
The difference between connections can be also written as (\ref{connections}).
For $A,B$ depending on $\phi, X$, the above expression can be expanded
\begin{eqnarray}
 \mathcal K\ud{\al}{\mu\nu} &=&
+ (\log A)_{,\phi} \left\{\phi_{,(\mu}\delta_{\nu)}^\al - B \ga^2 \phi^{,\al}\phi_{,\mu}\phi_{,\nu} - \half {A} \ga^2\phi^{,\al} g_{\mu\nu}\right\}
\nonumber \\ &&
+ (\log A)_{,X}\left\{-\phi^{,\sigma}\phi_{;\sigma(\mu}\delta_{\nu)}^\al + B \ga^2 \phi^{,\al}\phi^{,\sigma}\phi_{;\sigma(\mu}\phi_{,\nu)}
+\half\left[\phi_{,\sigma}\phi^{;\sigma\al} - B \ga^2\phi^{,\al}\enangle{\Phi}\right]g_{\mu\nu} \right\}
\nonumber \\ &&
+  B \ga^2 \phi^{,\al}\phi_{;\mu\nu}
+ \half B_{,\phi} \ga^2 \phi^{,\al}\phi_{,\mu}\phi_{,\nu} 
-B_{,X} \ga^2\phi^{,\al}\phi^{,\sigma}\phi_{;\sigma(\mu}\phi_{,\nu)}
+ \frac{B_{,X}}{2A}\phi_{,\mu}\phi_{,\nu}\lb \phi_{,\sigma}\phi^{;\sigma\al}
- B \ga^2\phi^{,\al}\enangle{\Phi}\rb\,.
\label{connectionlongX}
\end{eqnarray}

\subsection{Non-Horndeski Terms}\label{beyondHorndeski}

Starting with the Jordan frame in the form (\ref{eq:EH_Kfinal}) and plugging the $X$ dependent terms from the barred connection (\ref{connectionlongX}), the higher order part of the action is given by $ S_J[g_{\mu\nu},\phi] = \int  d^4 x\frac{\sqrt{-g}}{16\pi G}\Lag_{\rm disf}$ with 
\begin{eqnarray}
\Lag_{\rm disf} &=&
 \sqrt{A}\sqrt{A-2BX}\,R -\frac{ \sqrt{A}B}{\sqrt{A-2BX}}\enangle{R_{\mu\nu}}
+\frac{\sqrt{A} B^2 \left(\left\langle \Phi ^2\right\rangle -\langle \Phi \rangle  [\Phi ]\right)}{(A-2 B X)^{3/2}} 
-\frac{A_{,X} B \left(\left\langle \Phi ^2\right\rangle -\langle \Phi \rangle  [\Phi ]\right) (A-5 B X)}{\sqrt{A} (A-2 B X)^{3/2}}
 \nonumber \\ &&
+B_{,X} \left(\frac{\sqrt{A} B \left(\left\langle \Phi ^2\right\rangle -\langle \Phi \rangle  [\Phi ]\right) X}{(A-2 B X)^{3/2}}-\frac{A_{,X} (A-3 B X) \left(\langle \Phi \rangle ^2+2 \left\langle \Phi ^2\right\rangle  X\right)}{\sqrt{A} (A-2 B X)^{3/2}}\right)
 \nonumber \\ &&
+\frac{A_{,X}{}^2 \left(3 A^2 \left\langle \Phi ^2\right\rangle +8 B^2 X \left(\langle \Phi \rangle ^2+2 \left\langle \Phi ^2\right\rangle  X\right)-3 A B \left(\langle \Phi \rangle ^2+4 \left\langle \Phi ^2\right\rangle  X\right)\right)}{2 A^{3/2} (A-2 B X)^{3/2}}
 %\nonumber \\ &&
 + \Lag_3, \, \Lag_2 \text{ terms } \propto A_{,\phi},B_{,\phi}\,.
\end{eqnarray}
It is important to remember that the above expression only contains non-Horndeski terms.
Terms involving one instance of $A_{,\phi},B_{,\phi}$ can at most depend on $\enangle{\Phi}$, $[\Phi]$, and therefore contribute to $\Lag_3$ (\ref{LH3}). Terms involving two instances of $A_{,\phi},B_{,\phi}$ do not contain second derivatives, and therefore belong to $\Lag_2$ (\ref{LH2}). These terms have been indicated schematically in the last line.

\section{Frame Transformation for Special Disformal Mappings} \label{app:special_disf}

Let us now explore the transformation rules for gravitational theories under special disformal relations (\ref{eq:disf_special}).
Let us first consider the transformations of the Einstein-Hilbert and the Horndeski Lagrangians for the purely disformal case. Then the Gauss-Bonnet term will be presented in both the purely conformal and purely disformal cases.

\subsection{Einstein-Hilbert Lagrangian}

Let us consider a normalized, pure disformal relation
\begin{equation}\label{sweetbarg}
 \bar g_{\mu\nu} = g_{\mu\nu} + \pi_{,\mu}\pi_{,\nu}\,,
\quad \bar g^{\mu\nu} = g^{\mu\nu} - \ga_{0}^2 \pi^{,\mu}\pi^{,\nu}\,,
\end{equation}
with $\ga_{0}^2=\frac{1}{1-2X_\pi}$, $X_\pi = -\half \pi_{,\mu}\pi^{,\mu}$, and where the free function has been absorbed by a field redefinition $\pi = \int B(\phi)d\phi$. This simple form suffices to relate DBI Galileons to disformally coupled theories \cite{Zumalacarregui:2012us}. The field dependence will be restored in the final result.

%normalized special disformal geometry
The connection (\ref{connectionlongX}) and curvature tensor (\ref{riemmangen}) for the above relation are
\begin{eqnarray}\label{sweetkurva}
\mathcal{K}^{\al}_{\ph \mu\nu} = \bar g^{\al\la}\lp \pi_{,\la}\pi_{;\mu\nu}\rp = \ga_{0}^2 \pi^{,\al}\pi_{;\mu\nu}  \,. \label{simpleconn}
\end{eqnarray}
\begin{equation}\label{sweetriemann}
\bar R^\al_{\ph\bt\mu\nu} = \bar g^{\al\la}\lp R_{\la\bt\mu\nu} + \ga_{0}^2 \pi_{;\la[\mu}\pi_{;\nu]\bt} \rp  \,.
\end{equation}
Note that the form (\ref{sweetbarg}) has been assumed in the last expression to factor out the inverse barred metric (the first index can be then straightforwardly lowered: $ \bar R_{\al\bt\mu\nu}= R_{\al\bt\mu\nu} + \ga_{0}^2 \pi_{;\al[\mu}\pi_{;\nu]\bt}$). The Ricci tensor and scalar are given by
\begin{equation}
 \bar R_{\mu\nu} \equiv \bar R\ud{\la}{\mu\la\nu}
 = R_{\mu\nu} - \ga_{0}^2 R_{\al\mu\bt\nu}\pi^{,\al}\pi^{,\bt}
 + \ga_{0}^2 \left\{[\Pi]\pi_{;\mu\nu} - \pi_{;\nu\al}\pi\ud{;\al}{;\mu}\right\}
 - \ga_{0}^4\left\{ \enangle{\Pi}\pi_{;\mu\nu} - \pi^{,\al}\pi_{;\al\mu} \pi^{,\bt}\pi_{;\bt\mu} \right\}\,,
\end{equation}
\begin{equation}\label{riccisweet}
\bar R \equiv \bar g^{\mu\nu}\bar R\ud{\al}{\mu\al\nu} = R - 2\ga_{0}^2 \langle R_{\mu\nu} \rangle 
+ \ga_{0}^2\lp [\Pi]^2- [\Pi^2]\rp - 2\ga_{0}^4\lp [\Pi]\enangle{\Pi} - \enangle{\Pi^2}\rp\,.
\end{equation}

The transformed Einstein-Hilbert Lagrangian density can be obtained from Eq. (\ref{eq:EH_Kfinal})
\begin{eqnarray}
 \sqrt{-\bar g}\bar R &=& 
\sqrt{-g}\lp \frac{1}{\ga_{0}}R - \ga_{0} \enangle{R^{\mu\nu}}
- \ga_{0}^3\lp [\Pi]\enangle{\Pi} - \enangle{\Pi^2}\rp \rp
\label{LagGF0}
\\ &=& \sqrt{-g}\lp \frac{1}{\ga_{0}}R 
- \ga_{0} \lp [\Pi]^2- [\Pi^2]\rp + \nabla_\al \xi^{\al} \rp \,, \label{LagGF2} 
\end{eqnarray}
in terms of a total derivative which does not contribute to the bulk equations of motion.
\footnote{This can be shown by partial integration of the last term in (\ref{LagGF0})
\begin{equation}
 - \ga_{0}^3\lp [\Pi]\enangle{\Pi} - \enangle{\Pi^2}\rp =  (\nabla_\al \ga_{0}) ( \pi^{\al}\Box\pi - \pi^{;\al\bt}\pi_{,\bt})
= -\ga_{0}\lp [\Pi]^2 - [\Pi^2]\rp + \ga_{0} \enangle{R_{\mu\nu}} + \nabla_\al \xi^{\al} \,.
\end{equation}
with $\xi^\al= \ga_{0} (\pi^{\al}\Box\pi - \pi^{;\al\bt}\pi_{,\bt})$. The first equality uses the fact that $\nabla_{\mu}\ga_{0}=-\ga_{0}^3 \pi^{,\al}\pi_{;\al\mu}$, and the second follows after integration by parts and noting that $\pi^{,\bt}\nabla_{[\al}\nabla_{\bt]}\pi^{,\al}= 2 R_{\al\bt}\pi^{,\al}\pi^{,\bt}$.
}
The above Lagrangian has the right Horndeski form (\ref{LH4}) with $G_4=\ga_{0}^{-1}\equiv \sqrt{1-2X_\pi}$, $G_{4,X_\pi}=-\ga_{0}$ and $X_\pi = -\half \pi_{,\mu}\pi^{,\mu}$, therefore producing second order equations of motion.

It is possible to restore the field dependence in the disformal relation through a field redefinition $\pi =\int \sqrt{B(\phi)}d\phi$ in the Lagrangian density (\ref{LagGF2}). Then $\pi_{,\mu}=\sqrt{B}\phi_{,\mu}$, 
$\pi_{;\mu\nu} = \sqrt{B}\phi_{;\mu\nu} + \half \frac{B_{,\phi}}{\sqrt{B}}\phi_{,\mu}\phi_{,\nu}$ and the transformed Einstein-Hilbert term becomes
\begin{equation}
\sqrt{-\bar g}\bar R = \sqrt{-g}\left[ \frac{1}{\ga_b}R - B \ga_b \lp [\Phi]^2- [\Phi^2]\rp 
  + B_{,\phi}\ga_b \lp  2X[\Phi] + \enangle{\Phi}\rp \right] \,, \label{LagGFphi}
\end{equation}
with $\ga_b\equiv (1-2BX)^{-1/2}$, $X=-\half\phi_{,\mu}\phi^{,\mu}$. As these expressions contain no square roots of $B$, they are valid for negative values and recover the special case $B=-1$. Note that allowing $B$ to depend on $\phi$ adds lower order Horndeski terms, which are proportional to $B_{,\phi}$. These can be simplified by the addition of a total derivative%
\footnote{It is possible to remove the $f \enangle{\Phi}$ term in Eq. (\ref{LagGFphi}) by adding $\nabla_\al(g\phi^{,\al}) = g[\Phi] - g_{,X}\enangle{\Phi}-2Xg_{,\phi}$ with $g=\int f dX + s(\phi)$. Choosing $g=-\frac{B_{,\phi}}{\ga B}$ allows to obtain Eq. (\ref{L432_looking_nice}).}
\begin{equation}\label{L432_looking_nice}
\sqrt{-\bar g}\bar R = \sqrt{-g}\left[ \frac{1}{\ga_b}R - B \ga_b \lp [\Phi]^2- [\Phi^2]\rp 
+ \frac{B_{,\phi}}{\ga_b B}(\ga^2_d - 2)[\Phi] + 2X\lp\frac{B_{,\phi}}{\ga_b B}\rp_{,\phi} \right]   \,, \label{LagGFphi2}
\end{equation}
which corresponds to $G_4=\sqrt{1-2BX}$, $G_3=\frac{B_{,\phi}}{\ga_b B}(\ga^2_d - 2)$ and $G_2=2X\lp\frac{B_{,\phi}}{\ga_b B}\rp_{,\phi}$ in the original Horndenski form (\ref{LH2}-\ref{LH4}).

\subsection{Horndeski Lagrangian} \label{section:HorndeskiFr}

The transformation rules for the Horndeski Lagrangian (\ref{LH2}-\ref{LH5}) under special disformal maps are presented in Ref. \cite{Bettoni:2013cba}. In this section we derive the transformation rules for a normalized, pure disformal relation (\ref{sweetbarg}) in detail for $\Lag_2,\Lag_3$ and $\Lag_4$.  
The lowest order term is trivial to compute
\begin{equation}\label{disfL2}
\Lag_2=G_2(X_{\pi},\pi)
\to \frac{1}{\ga} G_2\lp \ga^2 X_{\pi},\pi \rp \equiv \frac{1}{\ga} \bar G_2\,,
\end{equation}
where the $\ga^{-1}$ factor arises from the barred volume element and a bar over a function means that the factor $\ga^2$ has been reabsorbed into the definition of the function $\bar G_i\equiv G_i(\ga^2 X_{\pi},\pi)$. Implicit dependence on $X_{\pi},\pi$ of the Horndeski functions will be assumed in the following. The next term is also simple to transform, noting that $\pi_{;\mu\nu} \to \bar \nabla_{\mu}\bar\nabla_{\nu} \pi = \ga^2\pi_{,\mu\nu}$
\begin{equation}
\Lag_3 = G_3[\Pi] 
\to \ga \bar G_{3}\lp[\Pi] - B\ga^2 \enangle{\Pi} \rp\,. \label{disfL3}
\end{equation}
See the footnote before Eq. (\ref{LagGFphi2}) on how to write cubic terms in Horndeski's form.

The quartic term is more complicated, but its Jordan Frame counterpart can be easily restored to a canonical Horndeski by noting that $ [\Pi]^2 - [\Pi^2]  \to \ga^4\left\{ [\Pi]^2 - [\Pi^2] -2\ga^2 \lp [\Pi]\enangle{\Pi} - \enangle{\Pi^2} \rp \right\} $, $G_{4,X_{\pi}}\to \bar G_{4,\bar X_{\pi}} = \bar G_{4,X_{\pi}} \lp \partial \bar X_{\pi} / \partial X_{\pi}\rp^{-1} = \ga^{-4}\bar G_{4,X_{\pi}}$ and following the same considerations used to transform (\ref{LagGF0}) into (\ref{LagGF2})
\begin{eqnarray}
 \Lag_4 =G_4 R + G_{4,X_{\pi}}\lp [\Pi]^2 - [\Pi^2] \rp 
 \to \frac{\bar G_4}{\ga}R + \lp \frac{\bar G_4}{\ga}\!\rp_{\!,X_{\pi}}\! \lp [\Pi]^2 - [\Pi^2]\rp 
 + 2\ga \bar G_{4,\pi} \lp\enangle{\Pi} + 2X_{\pi} [\Pi] \rp \,.
\end{eqnarray}
It can be seen that on top of a redefinition $G_4\to \frac{G_4}{\ga}$, if $G_{4,\pi}\neq 0$ a part of the Lagrangian is projected onto the lower order contribution $\Lag_3$ (last term).

\subsection{Gauss-Bonnet Term}\label{app:gauss_bonnet}

Besides the Ricci scalar present in the Einsten-Hilbert action, Lovelock's theorem allows for higher curvature terms whose variation gives second order equations of motion \cite{Lovelock:1971yv}. The following is the Gauss-Bonnet (GB) term, which does not contribute to the equations of motion in four dimensions. In this section we will compute the transformed GB term 
\begin{equation}\label{GBdef}
  \bar \mathcal G = \bar R^2 - 4 \bar R_{\mu\nu} \bar R^{\mu\nu} + \bar R_{\mu\nu\al\bt} \bar R^{\mu\nu\al\bt}\,
\end{equation}
for a special conformal and a normalized special disformal mapping.
Note that these results are essentially different from the projection of the bulk GB term into a codimension one submanifold, which is the usual approach in brane-world gravity \cite{Davis:2002gn}.

\subsubsection{Pure Conformal Relation}

Under a conformal transformation of the metric 
\begin{equation}
\bar g_{\mu\nu} = \Omega^2(\phi) g_{\mu\nu}\,,
\end{equation}
one finds the following transformation of the quadratic contractions
\begin{equation}\label{R2term}
 \bar R^2 = \Omega^{-4}\left[R^2  - 12 R\, \Omega^{-1} (\Box\Omega) + 36 \,\Omega^{-2}(\Box\Omega)^2 \right] \,,
\end{equation}
\begin{eqnarray}\label{Rmunu2term}
\bar R_{\mu\nu} \bar R^{\mu\nu} &\! = &\! \Omega^{-4}\left[ R_{\mu\nu} R^{\mu\nu} - 2 \,\Omega^{-1}\Big(2 R_{\mu\nu} \Omega^{;\mu\nu} + R \,\Box\Omega\Big) \right. 
\nonumber\\
&\!+&\!  \Omega^{-2}\Big(8 R_{\mu\nu}\, \Omega^{,\mu}\Omega^{,\nu} - 2 R\, (\partial\Omega)^2 + 4\Omega_{;\mu\nu}\Omega^{;\mu\nu} + 8 (\Box\Omega)^2 \Big) 
\nonumber\\
&\!-&\! \left. \Omega^{-3}\Big(4 \Omega_{;\mu\nu}\Omega^{,\mu}\Omega^{,\nu} - (\Box\Omega) (\partial\Omega)^2 \Big) + 12 \,\Omega^{-4} (\partial\Omega)^2\right] \,,
\end{eqnarray}
\begin{eqnarray}\label{Rmunurhosig2term}
\bar R_{\mu\nu\rho\sigma} \bar R^{\mu\nu\rho\sigma} &\! = &\! \Omega^{-4} \left[ R_{\mu\nu\rho\sigma} R^{\mu\nu\rho\sigma} - 8 \Omega^{-1} R_{\mu\nu} \Omega^{;\mu\nu} \right. 
\nonumber\\
&\!+&\!  4 \,\Omega^{-2}\Big( (\Box\Omega)^2 + 2 \Omega_{;\mu\nu}\Omega^{;\mu\nu} - R \,(\partial\Omega)^2 + 4 R_{\mu\nu}\, \Omega^{,\mu}\Omega^{,\nu}\Big) 
\nonumber\\
&\!+&\! \left. 8\, \Omega^{-3}\Big( (\Box\Omega)(\partial\Omega)^2 - 4 \,\Omega_{;\mu\nu}\Omega^{,\mu}\Omega^{,\nu} \Big) + 24 \,\Omega^{-4} (\partial\Omega)^2\right] \,,
\end{eqnarray}
and the transformed Gauss-Bonnet term reads
\begin{eqnarray}\label{GBterm}
\bar \mathcal G &\! = &\! \Omega^{-4}\left[ \mathcal G + 4\,\Omega^{-1} \Big( 2\,R_{\mu\nu} \Omega^{;\mu\nu} - R\,\Box\Omega\Big)  \right. 
\nonumber\\
&\!+&\!  2\,\Omega^{-2}\Big( 4\left((\Box\Omega)^2 - \Omega_{;\mu\nu}\Omega^{;\mu\nu} \right) - 8\,R_{\mu\nu}\, \Omega^{,\mu}\Omega^{,\nu} + 2 R\,(\partial\Omega)^2\Big) 
\nonumber\\
&\!+&\! \left. 8\,\Omega^{-3}\Big( 4\,\Omega_{;\mu\nu}\Omega^{,\mu}\Omega^{,\nu} - (\Box\Omega)(\partial\Omega)^2 \Big) - 24 \,\Omega^{-4} (\partial\Omega)^2\right] \,.
\end{eqnarray}

The GB action becomes, after integrating by parts (e.g.  terms $\nabla^\mu ( \Omega^{-7} \Omega_{,\mu} \Omega_{,\nu}\Omega^{,\nu})$  or  $\nabla^\mu (\Omega^{-1} R_{\mu\nu} \Omega^{,\nu})$),
\begin{equation}\label{GBaction}
 \int d^4x\,\sqrt{\bar g}\,\bar \mathcal G = \int d^4x\,\sqrt{g}\,\lp \mathcal G + \Delta \Lag_H\rp  \,,
\end{equation}
where the additional terms $\Delta \Lag_H$ can be expressed in Horndeski form (\ref{LH2}-\ref{LH5}) with
\begin{eqnarray}
\Delta G_2(\phi, X) &\! = &\!  - 176 \,\Omega^{-4} \,X_\Omega \,, \nonumber\\
\Delta G_3(\phi, X) &\! = &\!  - 48 \,\Omega^{-3} \,X_\Omega \,, \nonumber\\
\Delta G_4(\phi, X) &\! = &\!  8 \,\Omega^{-2} \,X_\Omega \,, \nonumber\\
\Delta G_5(\phi, X) &\! = &\!  - 8\, \Omega^{-1} \,,
\end{eqnarray}
and $\Omega=\Omega(\phi)$, $X_\Omega = -\frac{1}{2}(\partial\Omega)^2$.
Therefore, one concludes that adding the Gauss-Bonnet term to the Horndeski action does not change the structure of the theory under a purely conformal transformation, it merely changes the functions $G_i(\Omega(\phi), X)$.

\subsubsection{Normalized Pure Disformal Relation}

We will now compute the transformation rules for the Gauss-Bonnet term under a map given by a normalized, pure disformal relation (\ref{sweetbarg}). The $\bar R^2$ term follows from (\ref{riccisweet}), while the other terms read 
\begin{eqnarray}
\bar R_{\mu\nu} \bar R^{\mu\nu} &=& [R_{\mu\nu}^2] 
- 2\ga^2 \lp \langle R_{\mu\nu}^2\rangle - \langle R_{\mu\nu}R^{\al\mu\bt\nu}\rangle\rp
+ \ga^4 \lp \enaangle{ R_{\al\mu\bt\nu}R^{\ga\mu\da\nu} }+\enangle{R_{\mu\nu}}^2 \rp
\nonumber \\ && 
+2\ga^2\Big\{[\Pi][R_{\mu\nu}\Pi] - [\Pi R_{\mu\nu}\Pi]
+ \ga^2 \big( \enaangle{R_{\mu\al\nu\bt}\Pi^{\bt\ga}\Pi\du{\ga}{\bt}} + \enangle{\Pi R_{\mu\nu}\Pi } + 2\enangle{R \Pi^2}
-\enangle{R_{\al\mu\bt\nu}\Pi^{\al\bt}}[\Pi] 
\nonumber \\ && 
- 2\enangle{R_{\mu\nu}\Pi}[\Pi] - \enangle{\Pi}[R_{\mu\nu}\Pi] \big) 
+ \ga^4\big( \enaangle{\Pi^{\mu\al}R_{\la\al\sigma\bt}\Pi^{\bt\nu}} 
- \enangle{R_{\al\mu\bt\nu}\Pi^{\al\bt}}\enangle{\Pi} 
+ \enangle{R_{\mu\nu}}\lp \enangle{\Pi^2} -\enangle{\Pi}[\Pi]\rp 
\big)\Big\}
\nonumber \\ && 
+\ga^4 \Big\{ [\Pi^4] - 2[\Pi][\Pi^3] + [\Pi]^2[\Pi^2]
+ 2 \ga^2 \big( [\Pi^3]\enangle{\Pi} -[\Pi]^2\enangle{\Pi^2} 
-[\Pi][\Pi^2]\enangle{\Pi} + 3[\Pi]\enangle{\Pi^3} - 2\enangle{\Pi^4}\big)
\nonumber \\ && \phantom{+\ga^4}
+\ga^4\big(  \enangle{\Pi}^2([\Pi]^2 +[\Pi]^2) -2\enangle{\Pi}\enangle{\Pi^3}
-2[\Pi]\enangle{\Pi}\enangle{\Pi^2} + 2\enangle{\Pi^2}^2 \big)
\Big\} \label{GB2} 
\,,
\end{eqnarray}
\begin{eqnarray}
\bar R_{\al\bt\mu\nu} \bar R^{\al\bt\mu\nu} &=& [\![R_{\al\bt\mu\nu}^2]\!] 
- 4\ga^2 \enangle{R_{\mu\al\bt\ga}R^{\nu\al\bt\ga}} + 4\ga^4 \enaangle{R_{\al\mu\bt\nu}R^{\al\la\bt\sigma}}
\nonumber \\ &&
 + 4\ga^2\Big([\![\Pi^{\al\ga}R_{\al\bt\ga\da}\Pi^{\bt\da}]\!] +4\ga^2 \enangle{\Pi^{\al\bt}R_{\mu\al\bt\ga}\Pi^{\ga\nu}} 
+ 2\ga^4 \left\{\enangle{\Pi}\enangle{R_{\al\mu\bt\nu}\Pi^{\al\bt}} - \enaangle{\Pi^{\al\la} R_{\mu\al\nu\bt}\Pi^{\bt\sigma} }
\right\}\Big)
\nonumber \\ && + \ga^4 \Big\{ 2\lp [\Pi^2]^2 - [\Pi^4]\rp - 8\ga^2 \big( \enangle{\Pi^2}[\Pi^2]-\enangle{\Pi^4}\big)
+ 4\ga^4\big( \enangle{\Pi^2}^2 - 2\enangle{\Pi^3}\enangle{\Pi} + \enangle{\Pi}^2[\Pi^2] \big) \Big\}\,, \label{GB3}
\end{eqnarray}
The total result is
\begin{eqnarray}
\bar{\mathcal G} &=& R^2-4 [R_{\al\bt}^2]+[\![R_{\al\bt\ga\delta}R^{\al\bt\ga\delta}]\!]
-4 \gamma ^2 \Big\{ \enangle{R_{\al\bt\ga\mu}R^{\al\bt\ga\nu}} -2 \enangle{R_{\al\mu} R^{\al\nu}}
-2 \enangle{ R_{\al\mu\bt\nu}R^{\al\bt}} +R\  \enangle{ R_{\mu\nu}} \Big\} 
\nonumber \\ & &
+ \gamma ^2 \Big\{2 R [\Pi ]^2 -2 R [\Pi ^2]
+ 8 [\Pi  R_{\al\bt} \Pi ]
+4 [\![\Pi ^{\al\ga}R_{\al\bt\ga\delta}\Pi ^{\bt\delta}]\!]
-8 [\Pi ] [R_{\al\bt}\Pi]\Big\}
\nonumber \\ & &
-4  \gamma ^4\Big\{ 2 \enaangle{R_{\mu\al\nu\bt}\Pi ^{\al \ga}\Pi_\ga^\bt}
-4 \enangle{ \Pi ^{\al \bt}R_{\mu \al\bt\ga}\Pi^{\ga\nu} }
+4 \enangle{ R_{\al\bt}\cdot \Pi ^2} -R\enangle{ \Pi ^2 }
+2 \enangle{ \Pi R_{\al\bt} \Pi }
\nonumber \\ & &
-2   [\Pi ] \enangle{ R_{\al\mu\bt\nu}\Pi^{\al\bt}}
-4 [\Pi ] \enangle{ R_{\mu\al}\Pi } + R [\Pi ] \enangle{ \Pi}
+ \enangle{ R_{\mu\nu} }\big([\Pi ]^2  - [\Pi ^2] \big)
-2 \enangle{\Pi}  [R_{\al\bt}\Pi] \Big\}
\nonumber \\ & &
+\gamma ^4\Big\{[\Pi ]^4-6 [\Pi ]^2 \left[\Pi ^2\right]+3 \left[\Pi ^2\right]^2+8 [\Pi ] \left[\Pi ^3\right]-6 \left[\Pi ^4\right]\Big\} 
\nonumber \\ & &
+4  \gamma ^6 \Big\{ 6 \left\langle \Pi ^4\right\rangle -6 \left\langle \Pi ^3\right\rangle  [\Pi ]
+3 \left\langle \Pi ^2\right\rangle  [\Pi ]^2-\langle \Pi \rangle  [\Pi ]^3
-3 \left\langle \Pi ^2\right\rangle  \left[\Pi ^2\right]+3 \langle \Pi \rangle  [\Pi ] \left[\Pi ^2\right]-2 \langle \Pi \rangle  \left[\Pi ^3\right]\Big\} \label{GBdisf}
\end{eqnarray}
Here the terms arising from $R\cdot R$, $R\cdot\Pi$ and $\Pi\cdot\Pi$ correspond to the lines 1,2-4,5-6. The first three terms are just the Gauss-Bonnet term of the unbarred metric.

A theory whose Lagrangian density includes a $\int d^4 x\sqrt{-\bar g}\bar \mc G$ term of the above form does not belong to the Horndeski Lagrangian. This follows from the presence of terms terms with up to four contractions of the second derivatives of the scalar field $\pi$ in the transformed Gauss-Bonnet term (\ref{GBdisf}). However, we conjecture that the equations of motion for such a theory will be second order through the existence of implicit constraints (cf. section \ref{sect:frame_Jordan}), as the variation with respect to the metric would involve the Jacobian determinant and the higher order terms introduced by the disformal transformation would not be present in the original frame. The effects of the Gauss-Bonnet term in disformally coupled theories will be analyzed elsewhere.

\bibliographystyle{h-physrev}
\bibliography{disformal}

\end{document}